\documentclass[iop]{emulateapj}
\usepackage{graphicx}
\usepackage{epstopdf}
\usepackage{color}

\newcommand{\msun}{$\rm M_{\sun}$}
\newcommand{\rsun}{$\rm R_{\sun}$}
\newcommand{\rearth}{${\rm R_{\earth}} $}
\newcommand{\mearth}{$\rm M_{\earth}$}

\newcommand{\degree}{$^\circ$}

\newcommand{\first}{$1^{st}$}
\newcommand{\second}{$2^{nd}$}
\newcommand{\third}{$3^{rd}$}
\newcommand{\fourth}{$4^{th}$}

\definecolor{orange}{rgb}{.8,0.4,0}
\definecolor{darkred}{rgb}{.7,0.0,0}

\newcommand{\edit}[1]{{#1}}

\begin{document}

\shorttitle{}
\shortauthors{Berta et al.}
\title{Transit Detection in the MEarth Survey of Nearby M Dwarfs: \\ Bridging the Clean-First, Search-Later Divide}
\author{Zachory~K.~Berta\altaffilmark{1}, Jonathan~Irwin\altaffilmark{1}, David~Charbonneau\altaffilmark{1}, Christopher~J.~Burke\altaffilmark{2}, Emilio~E.~Falco\altaffilmark{3}}
\email{zberta@cfa.harvard.edu}
\altaffiltext{1}{Harvard-Smithsonian Center for Astrophysics, 60 Garden St., Cambridge, MA 02138, USA}
\altaffiltext{2}{NASA Ames Research Center, Moffett Field, CA 94035, USA ; SETI Institute, Mountain View, CA 94043, USA}
\altaffiltext{3}{Fred Lawrence Whipple Observatory, Smithsonian Astrophysical Observatory, 670 Mount Hopkins Road,
  Amado, AZ 85645, USA}

\begin{abstract}

In the effort to characterize the masses, radii, and atmospheres of potentially habitable exoplanets, there is an urgent need to find examples of such planets transiting nearby M dwarfs. The MEarth Project is an ongoing effort to do so, as a ground-based photometric survey designed to detect exoplanets as small as 2\rearth\ transiting mid-to-late M dwarfs within 33 pc of the Sun. Unfortunately, identifying transits of such planets in photometric monitoring is complicated both by the intrinsic stellar variability that is common among these stars and by the nocturnal cadence, atmospheric variations, and instrumental systematics that often plague Earth-bound observatories. Here we summarize the properties of MEarth data gathered so far, emphasizing the challenges they present for transit detection. We address these challenges with a new framework to detect shallow exoplanet transits in wiggly and irregularly-spaced light curves. In contrast to previous methods that clean trends from light curves before searching for transits, this framework assesses the significance of individual transits simultaneously while modeling variability, systematics, and the photometric quality of individual nights. Our Method for Including Starspots and Systematics in the Marginalized Probability of a Lone Eclipse (MISS MarPLE) uses a computationally efficient semi-Bayesian approach to explore the vast probability space spanned by the many parameters of this model, naturally incorporating the uncertainties in these parameters into its evaluation of candidate events. We show how to combine individual transits processed by MISS MarPLE into periodic transiting planet candidates and compare our results to the popular Box-fitting Least Squares (BLS) method with simulations. By applying MISS MarPLE to observations from the MEarth Project, we demonstrate the utility of this framework for robustly assessing the false alarm probability of transit signals in real data. 

\end{abstract}

\maketitle

\section{Introduction}
Observationally, nearby M dwarf stars offer both opportunities and challenges as exoplanet hosts. M dwarfs' low masses and small sizes accentuate the radial velocity wobble and eclipse depths of any planets that transit them. Their low luminosities result in habitable zones at much smaller orbital distances than for more luminous stars, so planets in M dwarf habitable zones are more likely to transit and will transit more frequently. These advantages aid the initial discovery \citep{nutzman.2008.dcgtshpod, blake.2008.nmudpstc} and the later detailed characterization \citep[e.g.][]{deming.2009.dctseuatsfjwst} of planets that could be small enough and cool enough to potentially host life. Mid-to-late M dwarfs offer a particularly compelling balance in that they have smaller statures than earlier-type stars but are still sufficiently bright to enable high precision followup studies, unlike later-type objects.

Exploiting this opportunity, the ground-based MEarth Project is using robotic, 40 cm telescopes to photometrically monitor nearby ($<33$ pc), mid-to-late M dwarfs. MEarth has been operating since 2008 with eight telescopes on Mt. Hopkins, AZ, and will soon include 8 additional telescopes in the Southern hemisphere. By design, MEarth intends to be sensitive to planets as small as 2\rearth\ and with periods as long as 20 days, reaching the habitable zones of these stars \citep[see][]{nutzman.2008.dcgtshpod}. \edit{ Like MEarth, several additional ground-based surveys are attempting to capitalize on the M dwarf advantage, including PTF/M-dwarfs \citep{law.2011.tewmbdstpam}, TRAPPIST \citep{jehin.2011.ttppst, bonfils.2012.utndgdwhvctwtp}, APACHE \citep{giacobbe.2012.ptspacsfwiaps}, and the WFCAM Transit Survey \citep{nefs.2012.fuembwts}.
}

\edit{
MEarth's first discovered transiting planet, the 1.6 day, 2.7 \rearth, 6.6 \mearth\ exoplanet GJ1214b  \citep{charbonneau.2009.stnls}, is far too hot for habitability. But as the first planet in this size range accessible to atmospheric characterization, GJ1214b has proven a useful laboratory for theoretical work \citep[e.g][]{miller-ricci.2010.nats1,rogers.2010.tpol1,nettelmann.2011.tesmts1,menou.2012.accg,miller-ricci-kempton.2012.ac1pc} and for observational studies, both from the ground \citep[e.g.][]{bean.2010.gtsse1,bean.2011.ontssgfema, carter.2011.tlcpxsts1, berta.2011.gsssvtsap, kundurthy.2011.ao1spesa, croll.2011.btss1smmwa,crossfield.2011.hdnts1,de-mooij.2012.ontos1wm} and from space \citep{desert.2011.oemasg,berta.2012.ftssgfwfchst}. Yet, its period is still very short. What are the prospects for finding planets with longer periods, potentially habitable planets?
}

\edit{

In light of the relative ease with which the space-based Kepler Mission can find transiting planets with periods longer than 100 days \citep[][]{batalha.2012.pcokafmd}, it is important to emphasize that 10-20 day habitable zone periods are long enough to pose significant detection challenges from the ground. Planets with these periods, even if geometrically aligned to transit, may offer only a single transit per season that can be observed from a single site, between weather losses and daytime gaps \citep[e.g.,][]{pepper.2005.stpss, von-braun.2009.owfpts}. The scarcity of transits poses a two-fold problem: multiple transits are often necessary to build up sufficient signal-to-noise for detection \citep[e.g.][]{bakos.2010.hsptbskf}, and multiple transits are, at some point, almost always necessary for determining a planet's period. 
}

\edit{

MEarth attempts to address the first challenge with a novel, automated ``real-time trigger'' mode of operation. This aids our ability to establish sufficient signal-to-noise to detect planet candidates from one or very few transits. While observing a target at low-cadence, MEarth can rapidly identify in-progress, marginally significant, single transit events from incoming observations. If an in-progress event crosses a low ($3\sigma$) threshold, MEarth can automatically trigger high-cadence followup to confirm the candidate event at higher confidence \citep[see][]{nutzman.2008.dcgtshpod, irwin.2009.mpsthsand}. If a transit is real, the triggered observations could magnify its significance from mediocre to ironclad, without having to wait to observe subsequent transits. If no transit is present, the triggered observations generally wash out the importance of the original downward outliers.
}

\edit{
The ability to confirm single events at high significance is crucial to MEarth's goal of finding long period planets. Our recent discovery of LSPM J1112+7626, a bright 0.4 + 0.3 \msun\ double-lined eclipsing binary in a 41 day orbital period \citep{irwin.2011.ljdmebfmts} highlights this point. We identified LSPM J1112+7626 from three exposures taken during a single primary eclipse. Due to the deep ($>10\%$) eclipses, we were confident the system was real. In parallel with continued photometric monitoring, we began radial velocity observations, and the combination of these two efforts ultimately established the binary's 41 day period. We envision the discovery of a long period planet to follow the same trajectory: a shallow ($1\%$) transit could be identified confidently using high-cadence observations from the real-time trigger, and follow-up scrutiny could be invested to measure the planet's period. This two-part strategy is the only way a ground-based survey like MEarth will have sufficient sensitivity to find planets with periods longer than $10$ days. }

\edit{
For this strategy to work, we need a robust method for accurately assessing the significance of individual transit events, both initially to trigger high-cadence observations of marginal events and later to assess whether an event is significant enough to warrant period-finding follow-up. This problem would be straightforward if MEarth's transit light curves exhibited no noise other than perfectly-behaved, uncorrelated, Gaussian, photon noise. This is not the case. MEarth light curves show astrophysical noise from the M dwarfs themselves, in the form of rotational modulation due to starspots or sporadic stellar flares. They show instrumental noise, such as that caused by pointing drifts, focus changes, and flat-fielding errors. They show extinction effects from Earth's dynamic atmosphere, some of which, as we discuss here, pose particularly pernicious problems for photometry of red stars. Often, these noise sources can mimic both the amplitude and the morphology of single planetary transits.
}

\edit{

To invest MEarth's follow-up efforts wisely, we need a conservative method for assessing the significance of a transiting planetary signal in the face of these complicated noise sources. This method needs to both (a) suppress, remove, or correct for stellar variability and systematics to increase sensitivity to shallower transits and (b) accurately propagate the uncertainties associated with this cleaning process into the significance assigned to the candidate signal. This method needs to be able to do so, even if only one or few transits are observed. It does not need to accurately determine the period of a signal; we postpone that endeavor for the eventual follow-up of statistically promising candidates.
}

\edit{

The exoplanet literature is teeming with well-established methods for cleaning  variability and systematics from transit survey light curves and for searching those cleaned light curves for periodic transit signals. However, to our knowledge, of those methods appropriate for ground-based observations none is sufficiently well suited to this challenge of estimating the significance of individual transits events. In this paper, we propose a new method, one that searches for transits simultaneously with a light curve cleaning process, so that the significance of candidates is marginalized over the cleaning's uncertainties.
}

\edit{

 If the rate of planet occurrence around mid-to-late M dwarfs rises sharply toward smaller planet sizes and long periods, as it does for FGK stars \citep{howard.2012.pow0ssfk}, then the development of even small improvements to our ability to detect shallow, rare transits could have a big payoff for MEarth. Additionally, the development of this method also provides a framework with which to estimate the ensemble sensitivity of the survey as a whole, thus enabling a statistical study from MEarth on the population of planets orbiting nearby mid-to-late M dwarfs. We intend to describe the results of such a study in a forthcoming work.
}

We begin by introducing the MEarth survey with a description of the observations we have gathered so far (Section \ref{s:observations}). After reviewing the light curve cleaning and transit detection techniques that have been described in the literature to date (Section {\ref{s:background}), we outline our new framework for MEarth, describing both how to estimate the significance of a single transit event and how to incorporate well-characterized single events into periodic planet candidates (Section \ref{s:missmarple}). We test this method with simulations of injected transits and demonstrate that the candidates generated by its application to the existing MEarth dataset have the statistical properties we would expect (Section \ref{s:results}). We conclude by suggesting other potential applications and improvements that could be made with this method (Sections \ref{s:future} and \ref{s:conclusions}). 

The reader should note that throughout this paper we use the terms ``eclipse'' and ``transit'' completely interchangeably, referring to a planet passing in front of its star as seen from Earth.

\section{Observations}\label{s:observations}
To frame the observational problem we hope to address, we summarize the properties of the photometric data gathered by the MEarth Project since beginning its full operation in 2008. For completeness, we reiterate some of the points described in the MEarth design strategy \citep{nutzman.2008.dcgtshpod}, emphasizing the qualitative features of the MEarth data that present particular challenges to our goal of detecting transits of habitable super-Earths.

\subsection{The Observatory}
Each of MEarth's eight telescopes is an f/9 40-cm Ritchey-Chr\'etien mounted on a German Equatorial mount. The telescopes are located in a single enclosure with a roll-off roof at the Fred Lawrence Whipple Observatory (FLWO) at Mount Hopkins, Arizona. They are robotically controlled and observe every clear night, except for instrument failures. Due to the summer monsoon in Arizona, we never observe during the month of August when FLWO is closed, and we rarely gather much useful data in July or September. 

Each telescope is equipped with a $2048\times2048$ CCD with a pixel scale of 0.76''/pixel, for a 26' field of view. Our target list contains 2,000 nearby M dwarfs \citep[selected from][]{lepine.2005.cnswapmlt0lc} that are spread all across the Northern sky ($\delta > 0$\degree), so they must be observed one-by-one, in a pointed fashion. The field of view is large enough to contain ample comparison stars for each MEarth target, with typically at least ten times as many photons available from comparisons as from the target. 

We use a custom 715 nm longpass filter, relying on the quantum efficiency of our back-illuminated e2v CCD42-40 detector to define the long-wavelength response of the system. Extending out to 1000 nm, the shape of this response resembles a combination of the Sloan $i+z$ filters \citep{fukugita.1996.sdsps}. The broad wavelength range of this filter was designed to maximize our photon flux from M dwarfs, but it introduces an important systematic effect into our photometry, as outlined in Section \ref{s:PWV}.

\subsection{Weather Monitoring}
MEarth continuously monitors the conditions on Mt. Hopkins with a suite of weather sensors. At ground level, we measure temperature, humidity, and wind speed, as well as rain and hail accumulation. We detect cloud cover with a wide-angle infrared sensor (a TPS-534 thermopile) that measures the sky brightness temperature at wavelengths $>5.5\micron$ \citep[see][]{clay.1998.cmsrs}. The primary purpose of this monitoring is to prevent damage to the telescopes by keeping the observatory closed during inclement weather, but the timeseries from this monitoring are also useful in later analysis for identifying weather-related systematics in our data.

\begin{figure*}[htbp]
\begin{center}
\includegraphics[width=\textwidth]{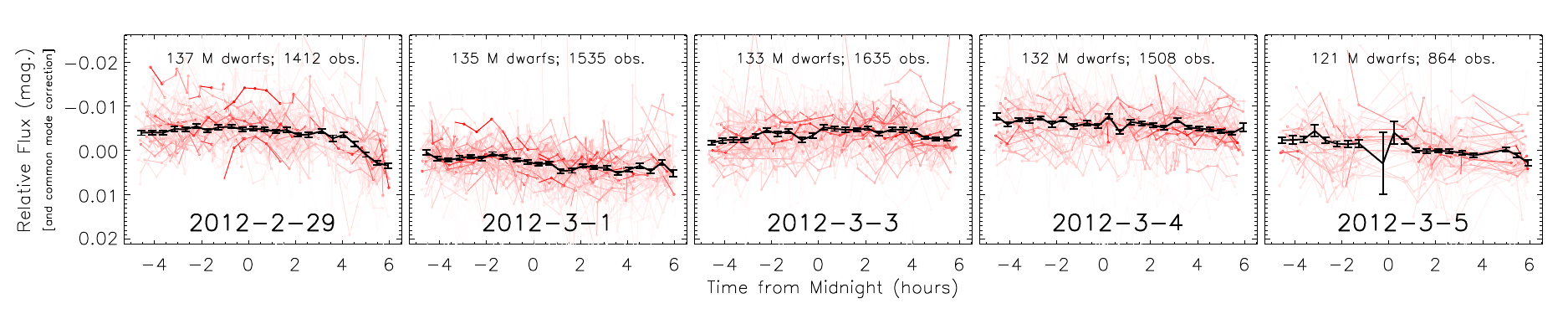}
\caption{One week of light curves of all M dwarf targets observed on all MEarth telescopes ({\em red lines}), along with a 30-minute median-binned estimate of their shared behavior ({\em black line}, with uncertainty estimates). This ``common mode'' shows significant variations both within and between nights. We attribute this phenomenon to variations in the precipitable water vapor above our telescopes changing the effective shape of our wide bandpass, effectively causing more extinction for red M dwarfs than for their bluer comparison stars. The common mode correlates strongly with measured humidity and sky temperature (see Figure \ref{f:cmcorrelations}). }
\label{f:cmdemo}
\end{center}
\end{figure*}

\begin{figure}[htbp]
\begin{center}
\includegraphics[width=\columnwidth]{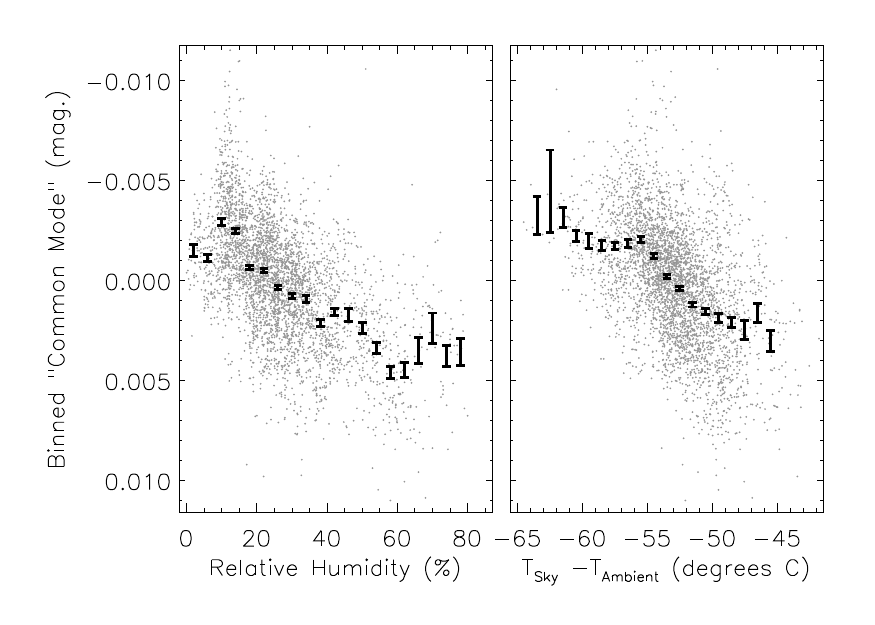}
\caption{The relationship between the shared ``common mode'' behavior in MEarth photometry and ground-level relative humidity ({\em left}) and the difference between sky and ambient temperatures ({\em right}), both very rough (and not necessarily linear) tracers of the total precipitable water vapor in the overlying column. For the entire 2011-2012 season, we show  each quantity averaged over independent half hour intervals (see Figure \ref{f:cmdemo}) as gray points. Error bars indicate the mean and its standard error for the common mode in subdivisions of humidity or sky temperature. In our bandpass, M dwarfs appear fainter when levels of precipitable water vapor are higher. }
\label{f:cmcorrelations}
\end{center}
\end{figure}

\subsection{Calibrations}\label{s:calibrations}

To go from raw images to reduced light curves, we follow the procedure and use modified code from \citet{irwin.2007.mpdplcp}. Here, we review those points in the process where calibration error could potentially lead to light curve systematics.

\subsubsection{Non-linearity}

The MEarth CCD's behave slightly non-linearly at all count levels, increasing up from a $1-2\%$ non-linearity at half of the detector full well up to $3-4\%$ near the onset of saturation. Because we often need to use comparison stars with different magnitudes than our target star, we must account for this non-linearity. When setting exposure times, we avoid surpassing 50\% of the detector's full well, and estimate a correction for the non-linearity using sets of daytime dome flats taken with different exposure times. With these measures in place, we see no evidence that non-linearity limits our photometric performance.

\subsubsection{Dark Current and Persistence}

We scale dark exposures taken at the end of each night to remove some of the CCD dark current. However, until 2011 we operated our Peltier-cooled detectors at -20\degree\ to -15\degree\ C, and at these warm temperatures they showed significant persistence. That is, images of bright stars would persist as excess localized dark current in subsequent images, slowly decaying with a half hour timescale below an initial 1\% fraction of the original fluence. This was a significant source of systematics: stars in incoming exposures could land on the same pixels as persistent ghost stars from previous exposures, thus gaining a hidden amount of flux that depended on how recently and strongly those pixels were illuminated. As this effect depends on the entire recent 2-dimensional illumination history of the detector, correcting for it would be extremely complicated.  \edit{We partially mitigated the persistence by ensuring different target stars were observed on different regions of the detector, but could not eliminate problems due to overlap with comparison stars or due to changes in cadence.} Updating our camera housings in 2011, we now operate at -30\degree\ C where the amplitude of the persistence is lower. We also adopt a detector preflash before each exposure; this increases the overall dark current but suppresses localized persistent images. Between the lower temperature and this preflash step, persistence no longer has a substantial effect on MEarth photometry.

\subsubsection{Flat-field Sensitivity Map}

We gather flat-fields at evening and morning twilight, typically 8 \edit{per telescope} per twilight, with empirically set exposure times estimated using the equations of \citet{tyson.1993.egttfwlfc}. To average out large-scale gradients in the illumination, we always take adjacent pairs of flats on opposite sides of the meridian, which has the effect of rotating the whole optical system relative to the sky (thanks to our German Equatorial mounts). Because our \edit{optical system} shows high levels of centrally concentrated scattered light \edit{($10-15\%$ of sky before 2011 and $<5\%$ after; see Table \ref{t:log})} that corrupts the large-scale structure in twilight flat exposures, we estimate the sensitivity in the detector plane in two steps. First, we estimate a small-scale sensitivity map that accounts for dust donuts and pixel-to-pixel variations in the detector sensitivity by filtering out large-scale structure from the combined twilight flats.  Second, we derive a large-scale map from dithered photometry of dense star fields to account for the non-uniform illumination across the field of view. Additionally, our camera's leaf shutter takes a finite time to open and close, resulting in a varying exposure time across the field of view (on a 1 second exposure, the amplitude of this effect is 5\%); we apply a shutter correction estimated from sets of twilight flats. Altogether, our flat-fielding procedure achieves a precision of 1\% across the entire detector. 

However, because we hope to perform photometry down to the level of 0.1\%, this 1\% knowledge of the sensitivity across the field is still imperfect and will inevitably be a source of systematics in our light curves. One unavoidable problem is that our German Equatorial mounts require the detector to flip 180\degree\ when crossing the meridian. In light curves, this causes offsets as large as 1\% between opposites sides of the meridian, as stars sample different regions of the large-scale sensitivity of the camera. Notably, the step-function morphology of this systematic can mimic a transit ingress or egress. In addition to this ``meridian flip'' problem, we achieve a blind RMS pointing accuracy of  60-120". To improve on this, at each pointing we take a short binned image and use its astrometric solution to nudge the telescope to the correct pointing before science exposures, with a random error typically of 1-2". This minimizes the impact of these pointing errors, but does not completely remove the problem of stars sampling different pixels on an imperfectly flat-fielded detector. 

\subsubsection{Differential Photometry}

We perform aperture photometry on all sources in the field of view. For each exposure, we derive a differential photometric correction from point sources in the field using an iterative, weighted, clipped fit that excludes variable stars from the comparison sample \citep[see][for details]{irwin.2007.mpdplcp}. We calculate a theoretical uncertainty estimate $\sigma_{\rm the}(t)$, in magnitudes\footnote{Technically, we convert from relative flux uncertainties into magnitude space as in \citet{naylor.2002.opcda2}, to which Eq.~\ref{e:sigma_the} is an accurate Taylor approximation.},  for each point:
\begin{equation}
\sigma_{\rm the}(t) = \frac{2.5}{\ln10} \times \frac{ \sqrt{N_{\gamma}  + \sigma_{\rm {sky}}^2 + \sigma_{\rm scint}^2 + \sigma_{\rm comp}^2}}{N_{\gamma}}
\label{e:sigma_the}
\end{equation}
where $N_{\gamma}$ is the number of photons from the source, $\sigma_{\rm sky}^2$ is an empirically determined sky noise estimate for the photometric aperture that includes read and dark noise, $\sigma_{\rm scint}^2$ is the anticipated scintillation noise \citep{young.1967.peavcrts}, and $\sigma_{\rm comp}^2$ accounts for the uncertainty in the comparison star solution. In some MEarth fields with very few comparisons, the $\sigma_{\rm comp}^2$ term can be a significant contribution to the overall uncertainty.

\subsubsection{Precipitable Water Vapor}\label{s:PWV}

A crucial assumption of this differential photometry procedure is that atmospheric or instrumental flux losses are exactly mirrored between target and comparison stars. Our wide 715-1000 nm bandpass overlaps strong telluric absorption features due to water vapor, so as the level of precipitable water vapor (PWV) changes in the column over our telescopes, their effective wavelength response will also change. Red stars will experience a larger share of this time-variable PWV-induced extinction than stars that are blue in this wavelength range. As a typical MEarth field consists of one very red target star (median target $r-J = 3.8$)\footnote{We take $r$ magnitudes from the Carlsberg Meridian survey \citep{evans.2002.cmtdss}, and $J$ magnitudes from 2MASS \citep{skrutskie.2006.ms2}.} amongst much bluer comparison stars (median comparison $r-J=1.3$), most MEarth M dwarfs exhibit systematic trends caused by this second-order extinction effect. This PWV problem has been noted before as a limitation for cool objects observed in the NIR \citep{bailer-jones.2003.lipmbd,blake.2008.nmudpstc}. Recently, \citet{blake.2011.maeugpsr} showed that GPS water vapor monitoring could be used to correct for the influence of PWV variations, improving both relative and absolute photometric accuracy of SDSS red star photometry.

While we do not have a GPS water vapor monitor, we can track the impact of PWV variations on MEarth photometry using the ensemble of observations we gather each night, observations of red stars in fields of blue comparisons. Figure \ref{f:cmdemo} shows all of the M dwarf light curves gathered by MEarth over one week, after applying basic differential photometry. These light curves (of different M dwarfs observed on different telescopes) move up and down in unison, reflecting water vapor changes in the atmosphere they all share. These trends correlate strongly with ground-level humidity and ambient sky temperature, which are rough tracers of PWV in the overlying column. As PWV variations within a night can mimic transit signals (e.g. the first panel of Figure \ref{f:cmdemo}), we must account for this effect when searching for planets. Fortunately, because these trends are shared among all our targets, we can estimate a ``common mode'' timeseries from the data themselves and use it to correct for these trends (see Section \ref{s:missmarple}).

\subsection{Science Observations}
The observations of our target M dwarfs are scheduled automatically using an ad hoc dynamic scheduling algorithm. This algorithm weights the observability of targets with the usefulness of the data to the survey as a whole, prioritizing gap-free cadences while minimizing slewing overheads. Each star is tied to a particular telescope, for ease of calibration and light curve production.

To inform this scheduling, we estimate masses, radii, and effective temperatures for all stars in the MEarth sample \citep{nutzman.2008.dcgtshpod}. Based on these estimates, we set the observational cadence to be sufficient to obtain two in-transit points from a mid-latitude transit of habitable zone planet. Because M dwarfs are dense stars, their transit durations are short (typically about 1 hour), requiring us to observe each star once every 20 minutes. 

Based on our estimated stellar radii, we set our exposure time for each star so that we will record as many photons as are necessary to \edit{for the transit of a 2\rearth\ planet to have a $3\sigma$ transit depth}. In cases where the required exposure time exceeds 2 minutes or would cause the peak counts in the star to exceed half of the detector full-well capacity, we split the observation into multiple sub-exposures. Stars requiring more than 7 minutes per pointing are never observed. If the time to reach 2\rearth\ is less than 60 seconds (i.e. bright, late M dwarfs), we artificially increase the exposure time. For the analyses presented in this paper, we combine all observations taken in a single pointing using scaled inverse-variance weighted means. 

The scheduler input list can be updated in real-time, allowing us to ``trigger'' high-cadence observations of the egress of interesting transit events that are detected in progress. By immediately gathering more observations in candidate transits, we can greatly magnify the significance of an initial $3\sigma$ detection or refute it entirely without having to wait for future transits. \edit{As currently implemented, the real-time trigger assesses the significance of ongoing transits after subtracting a fixed systematics model and harmonic variability model and inflating the theoretical error on the in-transit mean with an uncertainty estimate on the baseline out-of-transit level that is exponentially weighted toward recent observations. This practical estimator may eventually be replaced by the method explored in this paper.} Skimming each star with a minimal cadence and triggering on marginal candidates maximizes our overall efficiency and increases our sensitivity to long-period planets. 

\begin{figure*}[htbp]
\begin{center}
\includegraphics[width=\textwidth]{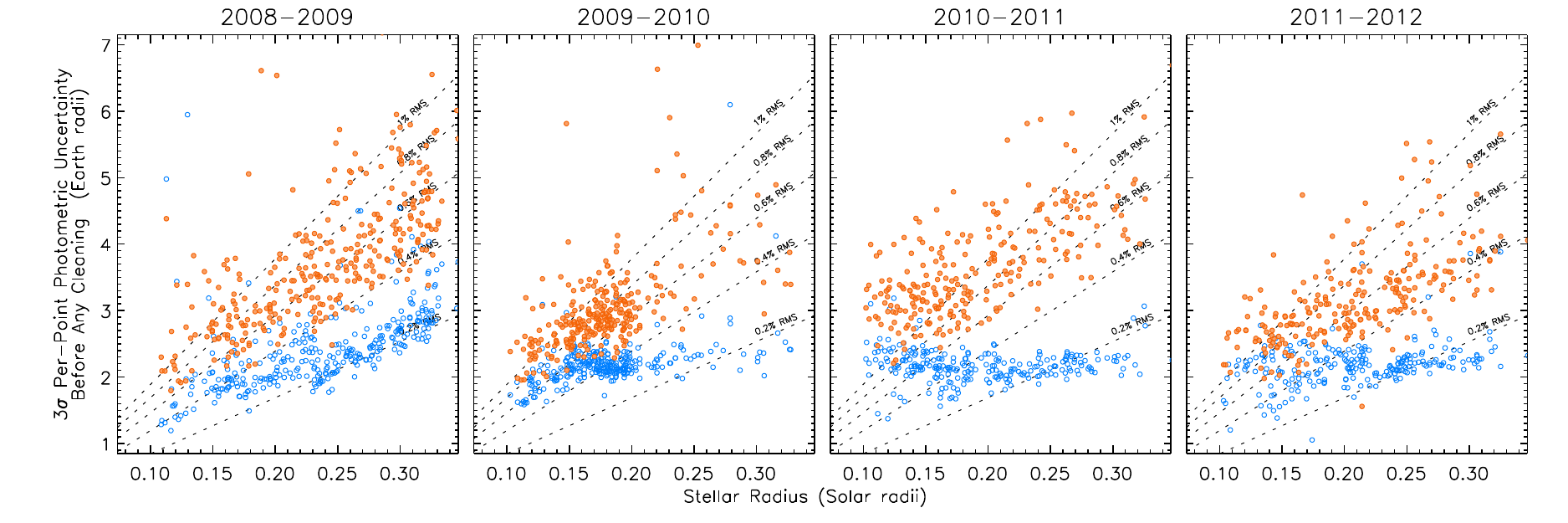}
\caption{The per-point RMS photometric uncertainty as predicted from a CCD noise model ({\em blue open circles}) and the RMS actually achieved in the raw differential photometry ({\em orange filled circles}) as a function of estimated stellar radius for all MEarth targets. We cast the photometric uncertainty for each M dwarf target in units of the \edit{planet radius corresponding to a $3\sigma$ transit depth, given the inferred stellar radius}; contours of constant RMS are shown for reference ({\em dashed lines}, \edit{equally spaced from 0.2 to 1\%}). \edit{Unlike wide-field surveys, we set exposure times individually for each target, minimizing the importance of apparent magnitude in these plots. The achieved RMS is shown} before any treatment of systematics or stellar variability; see Figure \ref{f:cleanrms} for comparison. One season of MEarth photometry is shown in each panel; Table \ref{t:log} explains the causes of many of the year-to-year variations.}
\label{f:rms}
\end{center}
\end{figure*}

\subsection{Morphological Description of the Light Curves}
A typical MEarth light curve for a target M dwarf contains roughly 1000 observations spanning one observational season. Most of the time, the 20 minute cadence is continuous \edit{within each night} for the time a star has a zenith distance $<60$\degree, but could be faster than this for up to several hours if a trigger occurred on the star. The cadence might also contain gaps \edit{within a night} due to passing clouds or if a trigger occurred on another star observed by the same telescope. On longer timescales, in addition to gaps \edit{for daylight}, light curves contain days- to months-long gaps from weather losses, instrumental failures, and scheduling conflicts (proximity to the Moon, other targets with higher priorities). 

One useful summary of the challenge MEarth light curves present is our achieved RMS scatter. For all M dwarfs observed in the past four years, we show in Figure \ref{f:rms} both the RMS predicted with our CCD noise model (Eq. \ref{e:sigma_the}) and the RMS actually achieved after basic differential photometry has been performed. To emphasize the implications for planet detection, we cast the RMS in terms of the size of planet that could be \edit{identified} at $3\sigma$ confidence in a single observation, given our stellar radius estimates.\footnote{Specifically, the vertical axis in Figure \ref{f:rms} is given by $\sqrt{3 \sigma} \times R_{\star}$, where $\sigma$ is the relative flux uncertainty in a single observation (either predicted or achieved) and $R_{\star}$ is the estimated stellar radius.} \edit{This choice of parameter space (over the more common RMS vs. apparent magnitude space) reflects two of MEarth's unique aspects. First, we know more about our target stars than most wide-field surveys, enabling this translation from RMS (and a corresponding detectable transit depth) into detectable planetary radius. Second, because we set exposure times individually for each star, our sensitivity to transits does not depend on stellar apparent magnitude.} The panels in Figure \ref{f:rms} show variations from year to year (see also Table \ref{t:log}), but all seasons of MEarth observations show significant gaps between the predicted and achieved noise. This indicates that stellar variability and systematics dominate over photon noise, highlighting the need for robust method to correct for these complicated noise sources in our search for transits. 

The source of excess scatter is sometimes known and sometimes unknown. By eye, some of the excess noise is clearly astrophysical, e.g. sinusoidal modulations from starspots rotating in and out of view or flares abruptly appearing then slowly decaying. Some is clearly instrumental, in that it can be associated with externally measured variables like position on the detector, weather parameters, or the behavior of other stars. Photometric outliers can often be associated with wind shake, where images exhibit broad and misshapen point spread functions. And lastly, some of the noise appears simply as unstructured excess scatter; this is caused either by astrophysical variability on timescales shorter than 20 minutes or by unidentified systematics.

\begin{table}
\caption{Evolution of MEarth Hardware/Software}
\begin{tabular}{r p{0.7\columnwidth}}
\hline\hline
Season & Notes \\ 
\hline
2008-2009 & Telescopes were operated purposely out-of-focus to minimize readout overheads, and exposure times were generally set to the maximum for a single (defocused) image, about 250,000 photons. The real-time trigger did not operate on-sky.\\
2009-2010 & After repeated focus mechanism failures resulted in many light curves experiencing large focus drifts, telescopes have been operated in or near focus since early in this season with the use of sub-exposures to avoid overexposure. Smaller stars were prioritized in the scheduling queue, as noticeable in Figure \ref{f:rms}. The real-time trigger began operating in November but was not always active due to development efforts. \\
2010-2011& In an attempt to remove systematics due to PWV, we operated during this year with a narrower filter ($715-895$nm, roughly $I_C$ in shape) designed to avoid strong telluric water features. Unfortunately, the interference cutoff of this filter was found to be sensitive to humidity and temperature, resulting in larger common mode variations and higher systematic noise in the light curves (see Figures \ref{f:rms} and \ref{f:rednoise_summary}). Scattered light was also more pronounced with these filters, spurring our multipart flat-fielding procedure.  The real-time trigger improved in its response time and its treatment of variability and the common mode. \\
2011-2012 & We returned to using the original MEarth 715 nm long-pass filter, but maintained the software improvements developed from the previous year. Dark flocking material affixed to the telescope baffles suppressed some of the scattered light. The real-time trigger operated normally for most of the season. \\
\end{tabular}
\label{t:log}
\end{table}

\section{Background}\label{s:background}

The problem of finding and assessing the significance of transiting exoplanet candidates in stellar photometry is an old one, and one that has already met many successful solutions. At their core, the majority of these solutions are variants of the matched filter Transit Detection Algorithm originally proposed by \citet{jenkins.1996.mfmgsdtepebad}, in which detection statistics are generated by matching light curves to families of templates consisting of periodic trains of transit-shaped pulses. The simplest and most intuitive of these methods is the Box-fitting Least Squares \citep[BLS;][]{kovacs.2002.baspt}, which models transits as simple boxcars in otherwise flat light curves. BLS identifies interesting candidates by folding individual photometric observations to trial periods, searching a grid of transit epochs and durations at each period, and picking the parameters that maximize the transit depth significance in a least-squares or $\chi^2$ sense. As discussed by \citet{aigrain.2004.csmpts}, many other matched filter methods \citep[][]{doyle.2000.oltipadsuptmwma,defay.2001.bmdpt, aigrain.2002.bdptmvgmbpsd,street.2003.sptfoc6,carpano.2003.dptpsvofci} are essentially generalizations of BLS.  

By assuming a flat out-of-transit light curve, BLS by itself can have a tendency to fold up any (non-planetary) time-correlated structures into seemingly significant candidates, when applied to real, wiggly light curves. As such, BLS is often paired with some sort of pre-search cleaning step to remove trends that could be caused either by instrumental effects or intrinsic stellar variability.

To deal with systematics, algorithms such as the Trend Filtering Algorithm \citep[TFA;][]{kovacs.2005.tfawvs} and the principal component analysis-like Systematics Removal method \citep[SysRem;][]{tamuz.2005.cselplc} were developed to remove trends that are present in multiple stars in a field and thus presumably not astrophysical. These algorithms use linear combinations of comparison star light curves to minimize the scatter in target stars. While these methods can remove trends without explicit knowledge of their causes, the trends do sometimes cluster into families that can be identified with physical processes \citep[e.g.][]{kim.2009.dtsavs}. Unfortunately, strategies \edit{like TFA} that work by constructing templates out of large numbers of field stars are of limited use for MEarth, with its small field of view and the substantial spectral type difference between our targets and comparisons. Methods that include known physical effects through linear models of externally measured variables \citep{bakos.2010.hsptbskf,ofir.2010.sadclcwsusep} are more helpful for MEarth-like data. 

Of course, systematics can also generally be minimized by improving various elements of the photometric reduction, observational strategy, or instrumentation. When it can be done at reasonable cost, this is always preferable to applying filtering methods after the fact, because filtering inevitably suppresses the desired signal in addition to the noise.

To clean stellar variability from light curves, many methods were developed in preparation for space transit surveys like CoRoT and Kepler, where precision photometry makes it a dominant concern \citep[e.g.][]{defay.2001.bmdpt,jenkins.2002.isvdttp,carpano.2003.dptpsvofci,aigrain.2004.ppp,regulo.2007.twardpt,bonomo.2008.msvdeptiptmhff}. These methods operate in the time, wavelet, or Fourier domains; many of them assume uniform photometric uncertainties and uniform cadence, as can realistically only be achieved from space. Running median filters \citep[e.g.][]{aigrain.2004.ppp} or piece-wise polynomial/spline fits \citep[e.g.][]{croll.2007.ls1sstmsp} have also proven effective for removing smooth variability from high S/N light curves. Ground-based surveys for planets in open clusters motivated new methods to remove large amplitude variability from light curves with diurnal gaps \citep{street.2003.sptfoc6,bramich.2005.sptf7, burke.2006.stepssilfswpoc1,aigrain.2007.mpsoyoc,miller.2008.mpstoc2}, often by fitting series of sinusoids or allowing slowly varying baselines. We refer the reader to reviews and comparisons of these methods by \citet{tingley.2003.ietdatc}, \citet{aigrain.2004.ppp}, and \citet{moutou.2005.cbtfptdarslc}. 

\edit{NASA's spaced-based Kepler Mission published the first Earth-sized planets \citep{fressin.2012.epok} and over 2,300 transiting planet candidates \citep{batalha.2012.pcokafmd} at the time of this writing. This success is thanks both to the design and stability of the spacecraft and to the sophistication with which the Kepler team accounts for its noise sources. To identify candidate transiting planets and assess their significance, Kepler employs a wavelet-based matched filter that is both optimal and efficient  \citep{jenkins.2002.isvdttp, tenenbaum.2012.dptsftqkmd}. The Kepler Pre-search Data Conditioning pipeline can also disentangle instrumental systematics from stellar variability using a linear model like the ones above paired with a Maximum A Posteriori approach employing empirical priors on the decorrelation coefficients to prevent over-fitting \citep[PDC-MAP;][]{smith.2012.kpdcbasec,stumpe.2012.kpdcaaecklc}. Unfortunately, due to the need for uniformly spaced data to run the wavelet filter, the applicability of the Kepler transit-search method is limited in ground-based observations. }

\edit{
The coupling of many of the above cleaning methods with the BLS search has proven extremely successful for wide-field surveys. Using these methods, surveys such as TrES \citep{alonso.2004.ttpbs}, HATNet \citep{bakos.2004.wmpwtepd}, XO \citep{mccullough.2005.pstepc}, WASP \citep{pollacco.2006.wpsc} and KELT \citep{siverd.2012.ksihispjctms,beatty.2012.kjtb8psbs} have made the ground-based detection of hot gas giants transiting Sun-like stars routine. Amidst these successes, why should we bother to develop new methods? 
}

\edit{
Most of the above cleaning methods that are suitable for use from the ground work by subtracting some optimized model for systematics and variability from a target light curve. Subtracting this model inevitably introduces some extra uncertainty to the light curve: a cleaned light curve cannot possibly be as reliable as a light curve that did not need to be cleaned in the first place. However, these methods generally do not include a route for propagating the uncertainty from this cleaning into the significance of candidate transits. When many transits will be folded into into a planet candidate, this is okay. It is sufficient to know the average effect the cleaning has on the light curve, for example, that global filtering with TFA suppresses transit depths by 20\% on average in HATNet \citep{bakos.2012.hgnfaiwt}. In contrast, when only a single transit is available, knowing the average effect is not enough. We need to know: to what extent can we say that any one given dip is a bona fide eclipse and not the result of over- or under-correction by the cleaning process?
}

\edit{
We need a method that escapes the clean-first, search-later dichotomy of many of the previous methods. If the cleaning and the search are a two step process, the search knows nothing about how the cleaning has suppressed or exaggerated the apparent significance of transit-like features, making establishing rigorous detection thresholds very difficult. We need a reliable way to include our uncertainty in the corrections we make for systematics and variability in our search for planets; one way to do this is to combine the steps together, allowing the search to know about all the complicating details that go into the cleaning. 
}

\edit{
Here, we present a new method to detect single transits and robustly assess their significance. With 10--20 day M dwarf habitable zone planets offering at most a handful of observable transits per season, the ability to identify promising candidates with one or very few transits is absolutely necessary to our success. Here, we present a method for folding single transits into phased planet candidates, but we do not focus extensively on the problem finding the true periods of candidate systems. Although the challenge MEarth faces is not as bad as for the most sparsely sampled light curves \citep[see][]{dupuy.2009.dtjlebsspsd,tingley.2011.stdwltbps,dzigan.2012.dtjegpy},  it will generally be extremely difficult to find accurate periods for 10--20 day planets from MEarth survey data alone. Rather, our goal is to be able to assess which candidates have high enough significance that they warrant the allocation of follow-up resources to, eventually, establish their periods and confirm their planetary nature. 
}

\section{Investigating a Single Eclipse: MISS MarPLE}\label{s:missmarple}

We start by assessing the significance of an individual transit event within the context of a single night of observations of a  star. We do so in the context of a parameterized, generative model for each target star light curve. This model contains parameters describing a simple box-shaped eclipse model, as well as parameters describing systematic effects plaguing the light curve and the star's intrinsic stellar variability. 

Of the many parameters in this model, the depth $D$ of a putative planetary eclipse is particularly important. We are interested in answering the following question: given a hypothetical lone planetary transit, with an epoch $p_{E}$ and duration $p_{T}$, what is the probability distribution of the planetary eclipse depth $D$ that the data imply? The integral of the normalized probability distribution $P(D | p_{E}, p_{T})$ over the range $D>0$ would provide a measure of the detection significance of the single eclipse. While $P(D)$ could generally take on any shape, we will approximate its shape to be Gaussian, so that we can completely characterize the distribution with two numbers, the maximum probability depth $\overline{D}$ and a width $\sigma$. In usual astronomical parlance, if $\overline{D}/\sigma > n$ then we have detected the eclipse ``at $n \sigma$.'' 

We want $P(D)$ to be conditional only on the parameters $p_{E}$ and $p_{T}$; it should be marginalized over all other parameters to account for the additional uncertainty that each of these add to the width of the distribution. That is, we want $P(D)$ to be the Marginalized Probability of a Lone Eclipse (MarPLE), whose Gaussian width we will refer to as $\sigma_{\rm MarPLE}$. In particular, because the transit depth could conceivably be quite correlated with the stellar variability or systematics parameters, marginalizing over these parameters will be crucial for a robust measure the eclipse depth uncertainty and thus the significance of the detection. To achieve this goal, we outline a Method to Include Starspots and Systematics in the Marginalized Probability of a Lone Eclipse (MISS MarPLE) below.

\subsection{The Model}\label{s:The Model}
At the core of MISS MarPLE is a model that attempts to describe every aspect of a single night of MEarth photometry of a single M dwarf. We use $d(t)$ to refer to  the ``data'' sampled at time $t$: the relative flux measurements of the target star after basic differential photometric corrections have been applied. The two main aspects of the model are an idealized, noiseless light curve $m(t)$ and the uncertainty associated with a data point at any given time $\sigma (t)$. This model is generative, in the sense that fake light curves created with this model aim to be statistically equivalent to real MEarth light curves. Even if the model is an incomplete description of $d(t)$, it will still be useful for estimating the significance of a given candidate by allowing us to fit for and marginalize over the model parameters.

Throughout the following sections, light curves such as $m(t)$ and $d(t)$ will be expressed in magnitudes, so that effects that are multiplicative in flux can be described as linear models. 

We write the model for the idealized, noiseless light curve as
\begin{equation}
m(t) =  S(t) + V(t) + P(t) 
\label{e:m(t)}
\end{equation}
where $S(t)$ models trends caused by instrumental {systematics}, $V(t)$ models the {variability} of the star in the absence of planetary transits, and $P(t)$ models the signal from a hypothetical transiting {planet}. 

\subsubsection{Systematics Model}
The $S(t)$ term in Eq.~\ref{e:m(t)} enables us to include systematic trends that show clear correlations with externally measured variables. We construct $S(t)$ as a linear combination of $N_{\rm sys}$ relevant external templates:
\begin{equation}
S(t) = \sum_{j=1}^{N_{\rm sys}}s_{j}E_{j}(t).
\label{e:S(t)}
\end{equation}
Here $E_{j}(t)$ represent timeseries of the external variables, sampled at the times as the photometric observations, and the $s_{j}$ are systematics coefficients. For MEarth, at a bare minimum, we include $N_{\rm sys}=6$ terms in this sum: the ``common mode,''  the ``meridian flip'', and the $x$ and $y$ pixel positions on either side of the meridian. 
\begin{description}
\item[$E_{\rm CM}(t)$] The common mode template is constructed from the ensemble of raw M dwarf light curves from all telescopes and accounts for photometric trends that are shared in all MEarth M dwarf photometry (due to PWV variations, see Figures \ref{f:cmdemo} and \ref{f:cmcorrelations}). The effect is stronger for redder stars; for MEarth targets, the best fit values of the coefficient $s_{\rm CM}$ correlates with stellar $r-J$ color.
\item[$E_{\rm merid}(t)$] To account for stars sampling different regions of the detector when observing at positive or negative hour angles with MEarth's German Equatorial mounts, we include a ``meridian flip'' template. This template is simply defined as 0 for observations taken in one orientation and 1 for observations in the other, thus allowing light curves on two sides of the meridian flip to have different baselines.\footnote{In practice, we also allow additional offsets corresponding to each time a camera is taken off of its telescope. This is implemented as a simple extension of the $E_{\rm merid}(t)$ term described here.}
\item[\textnormal{$E_{x,i}(t)$ and $E_{y,i}(t)$ for i=0,1}] The pixel position templates are simply the $x$ and $y$ centroids of the target star on the detector, with their medians subtracted. Two sets are required, one for each side of the meridian. Correlations with these templates could arise as pointing errors allow a star to drift over uncorrected small-scale features in the sensitivity of the detector (e.g., transient dust donuts). 

\end{description}
Additional external variables may also be used as systematics templates, such as FWHM or airmass. With MEarth, we find these variables are correlated with the photometry for only a few fields, and are usually excluded. 

\subsubsection{Variability Model}
The $V(t)$ term in Eq.~\ref{e:m(t)} describes the variability of the star throughout one night, independent of the presence of a transiting planet. Such variability includes fluctuations due to rotating spots (smoothly varying on the 0.1 to 100 day timescale of the star's rotation period) and flares (impulsively appearing, with a decay timescale typically of hours). The morphology of this variability can be quite complicated; we use a simplified model to capture its key features, writing
\begin{eqnarray}
V(t) =  &&v_{{\rm night}} +\nonumber \\
&&v_{\sin}\sin\left(\frac{2\pi t}{v_{P}}\right) +  v_{\cos}\cos\left(\frac{2\pi t}{v_{P}}\right)+\nonumber \\ 
&&\sum_{j=1}^{N_{\rm flares}} f_{j}(t).
\label{e:V(t)}
\end{eqnarray}
The first term $v_{\rm night}$ allows each night to have its own baseline flux level. By itself, this term can capture most of the variability from stars with long rotation periods, where the flux modulation from starspots smoothly varies over timescales much longer than one night. By fitting for a different $v_{\rm night}$ for each night we can piece together the variability of the star on timescales $>1$ day as a series of scaled step functions. The harmonic $v_{\sin}$ and $v_{\cos}$ terms capture variability with period of $v_P$ and become especially important for stars with shorter rotation periods. Because we fit a separate $v_{\rm night}$ for each night, there can be substantial degeneracy between the harmonic terms and the nightly offsets, especially for slowly rotating stars. We discuss this issue, as well as how we estimate $v_P$ in Section \ref{s:posterior}. Although we only include one harmonic of the fundamental period $v_P$ in these sinusoidal terms, additional harmonics could be included if the data warranted them. 

The final term in Eq. \ref{e:V(t)} includes contributions from $N_{\rm flares}$ hypothetical stellar flares $f_{j}(t)$ that may or may not be present within the night. \edit{Flares are suppressed in MEarth's relatively red bandpass, but not completely eliminated  \citep[see][]{tofflemire.2012.idfdceiw}. The main purpose of the flare term is to identify those nights of photometry that may be corrupted due to the presence of flares. While we could in principle model a night that contained both flares and a transit, we find this to be very difficult in practice, due to the morphological complexity flares sometimes exhibit \citep[e.g.][]{kowalski.2010.wlmds, schmidt.2012.pfaduiel}. Rather, on each night we model simple hypothetical flares as fast-rising and exponentially decaying, and perform a grid search over the start time and decay timescale. If any flares have amplitudes that are detected at $>4\sigma$, we excise that night of data from our planet search. These cuts dramatically reduces planetary false positives due to flaring activity (e.g. confusing the start of the flare with the egress of a transit), with the meager cost of ignoring 3.6\% of MEarth's observations. Because both planetary transits and flares are rare in MEarth data, and their overlap even moreso, the losses from this strategy are small.}

\subsubsection{Planetary Eclipse Model}
The last term in Eq. \ref{e:m(t)}, $P(t)$, includes the signal of a hypothetical transiting planet. We model transits as having infinitely short ingress/egress times and ignore the effects of limb-darkening on the host star, so transits appear as simple boxcars. In this section, we are interested only in assessing the significance of a single transit event falling within a single night, not a periodic train of transits. With these simplifications, a lone planetary eclipse signal is completely described by a transit epoch $p_{E}$, a transit duration $p_{T}$, and a transit depth $D$. The signal is then simply
\begin{equation}
P(t)= \left\{ \begin{array}{ll}
D & \mbox{if $|t - p_E| < p_T/2$}  \\
0 &\mbox{otherwise}
       \end{array} \right.\\ 
\end{equation}
This model includes only one eclipse event per night. We discuss combining these lone eclipses into periodic transit candidates in Section \ref{s:phasing}. 

\subsubsection{Photometric Uncertainty Model}

A crucial component of the model is $\sigma(t)$, the photometric uncertainty of each observation. Our theoretical uncertainty estimate for a given datapoint $\sigma_{\rm the}(t)$ is a lower limit on the true uncertainty. To express this fact, we introduce a noise rescaling parameter $r_{\sigma,w}$ such that 
\begin{equation}
\sigma(t) = r_{\sigma, w}\sigma_{\rm the}(t).
\label{e:rescaling}
\end{equation}
where $ r_{\sigma, w} \ge 1$. The subscript $w$ emphasizes that this is a {\em white} noise rescaling parameter that does not account for correlations between nearby data points. If left unmodelled, such {\em red} noise could substantially bias a transit's detection significance \citep{pont.2006.enptd}; we discuss a correction for red noise in \S\ref{s:rednoise}.

\subsection{The Posterior Probability}\label{s:posterior}
For a reasonable choice of parameters, the model in Eq. \ref{e:m(t)} could generate a fake light curve that would have most of the features of single night of a real MEarth light curve. But how do we pick a reasonable choice of parameters? In this section, we write down their probability distribution and show how to solve for its peak, which turns out to be a linear minimization process with slight iterative refinement. 

Considering a single night of observations, we write the shape of the probability distribution of these parameters as
\begin{equation}
P(\mathbb{M}|\mathbb{D}) \propto P(\mathbb{D}|\mathbb{M}) P(\mathbb{M}),
\label{e:posterior}
\end{equation}
where $P(\mathbb{M}|\mathbb{D})$ is the posterior probability of the model $\mathbb{M}$ given the data $\mathbb{D}$, $P(\mathbb{D}|\mathbb{M})$ is the likelihood of the data given the model, and $P(\mathbb{M})$ is the prior probability of the model. These functions describe probability density distributions that live in an $n$-dimensional hyperspace with as many dimensions as there are parameters in the model.

\subsubsection{The Likelihood = $P(\mathbb{D}|\mathbb{M})$}
We describe each of the $N_{\rm obs}$ photometric observations $d(t_i)$ within a particular night as being drawn from a Gaussian distribution centered on $m(t_i)$ and with a variance of $\sigma(t_i)^2$. Assuming the observations to be independent, the likelihood can be written as
\begin{eqnarray}
P(\mathbb{D}|\mathbb{M})= \prod_{i=1}^{N_{\rm obs}}\frac{1}{\sqrt{2\pi}\sigma(t_i)}\exp{\left[-\frac{1}{2}\left(\frac{d(t_i)- m(t_i)}{\sigma(t_i)}\right)^2\right]}. \nonumber
\end{eqnarray}
Taking the logarithm, substituting Eq. \ref{e:rescaling}, and defining
\begin{equation}
\chi^2 = \sum_{i=1}^{N_{\rm obs}}\left[\frac{d(t_i)- m(t_i)}{\sigma_{\rm the}(t_i)}\right]^2,
\end{equation}
we find that the (log) likelihood simplifies to
\begin{equation}
\ln P(\mathbb{D}|\mathbb{M}) = - N_{\rm obs} \ln r_{\sigma, w} - \frac{\chi^2}{2r_{\sigma, w}^2} + {\rm constant}
\label{e:loglikelihood}
\end{equation}
where we have only explicitly included terms that depend on the parameters of the model. For fixed $r_{\sigma, w}$, maximizing Eq. \ref{e:loglikelihood} is equivalent to minimizing the commonly used $\chi^2$ figure of merit. 

\subsubsection{The Prior = $P(\mathbb{M})$}\label{s:priors}
For any one star, a particular night of MEarth photometry may contain roughly as many light curve points as there are parameters in our model. As such, the likelihood $P(\mathbb{D}|\mathbb{M})$ from one night of data only very weakly constrains the parameter space. But of course, each night of MEarth observations is just one of many nights spanning an entire season, and we should use this season-long information when investigating a single night. To implement this holistic awareness of the context provided by a large pool of observations, we generate probability distributions for various parameters by looking at the whole season of data. We then apply them as priors $P(\mathbb{M})$ on the parameters for an individual night. 

By construction, the most important parameters of our model are linear parameters. The conditional likelihood of linear parameters (a slice through $P(\mathbb{D}|\mathbb{M})$ with other parameters fixed) has a Gaussian form. For marginalization, it proves quite useful for the priors to be conjugate to this shape -- that is, also take on a Gaussian form. Referring to these linear parameters with the vector $\mathbf{c}  =\{D, v_{\rm night}, v_{\sin}, v_{\cos}, s_{\rm CM}, s_{\rm merid}, s_{x,i}, s_{y,i}, s_{\rm other?}\}$, we parameterize the prior $P(\mathbb{M})$ as being proportional to a Gaussian distribution in $c_{j}$ that is centered on an expectation value $\overline{c_{j}}$ and with a variance of $\pi_{c_{j}}^2$. Multiplying the independent priors for the $N_{\rm coef}$ coefficients and defining
\begin{equation}
\Phi^2 = \sum_{j=1}^{N_{\rm coef}}\left(\frac{c_j - \overline{c_j}}{\pi_{c_j}}\right)^2 
\end{equation}
leads to a term in the prior that looks like
\begin{equation}
\ln P(\mathbb{M}) =  -\frac{1}{2}\Phi^2+ \cdots
\label{e:prior_linear}
\end{equation}
The similarity in form of $\Phi^2$ to $\chi^2$ is the reason that the use of conjugate Gaussian priors is often described along the lines of ``adding a prior as an extra data point in the $\chi^{2}$ sum,'' because the effect  is identical in the overall posterior. In this framework, the smaller values of $\pi_{c_j}$ provide tighter constraints on the parameter; we could express a flat, non-informative prior for a particular $c_{j}$ by choosing a large value of $\pi_{c_j}$.  We set $\pi_D = \infty$, giving a flat prior on the transit depth. \edit{Note that we allow negative transit depths (i.e. ``anti-transits'') to avoid skewing the null distribution of transit depths away from 0.}

For most of the remaining linear parameters, we take the values of $\overline{c_{j}}$ and $\pi_{c_{j}}$ directly from the results of a \edit{simultaneous fit to the star's entire season of observations.}  In this prior-generating season-long fit, we fit the season-long light curve with a modified version of Eq. \ref{e:m(t)} that excludes both the $f_j(t)$ term from flares and the $P(t)$ term from hypothetical planets. To immunize against these unmodelled flares and eclipses, we perform the fit with $4\sigma$ clipping. We prefer to explain as much of the long-term variability as possible with the harmonic terms, so we fit first including only these terms in $V(t)$. Then, fixing the values of $v_{\sin}$ and $v_{\cos}$, we fit again with one $v_{{\rm night}, j}$ free parameter for each night represented within the season. Thus, the values of $v_{{\rm night}, j}$ then represent the deviation of the nightly flux level from a baseline sinusoidal model. 

Now, for the single night flux baseline parameter $v_{\rm night}$, we set $\pi_{v_{\rm night}}$ equal to $1.48\times{\rm MAD}$ (median absolute deviation) of the ensemble of $v_{{\rm night}, j}$ values from the season fit. Stars that vary unpredictably from night to night will have a broad prior for $v_{\rm night}$, thus requiring more data within a night to determine its baseline level. Conversely, stars that remain constant from night to night or have variability that is well described by a sinusoid will have a very tight prior. 

To understand the impact of $\pi_{v_{\rm night}}$, imagine the following hypothetical scenario: a night in which MEarth gathered only one observation of a star, and that observation happened to fall in the middle of a transit with a 0.01 magnitude depth. With what significance could we detect this transit? If $\pi_{v_{\rm night}} = 0.01$ magnitudes, then the detection significance would be at most $1 \sigma$. But if $\pi_{v_{\rm night}} = 0.001$, then the transit could in principle be detected at high significance with only the single data point, provided the photon noise limit for the observation was sufficiently precise. 

We note that $s_{x, i}$ and $s_{y,i}$, the coefficients for the $x$ and $y$ pixel position templates, would not be expected to be constant throughout a season. These terms are designed to account for flat-fielding errors, which could easily change from week to week or month to month. As such, we do not take $c_{j}$ and $\pi_{j}$ from the season-wide fit for these parameters. Rather, we fix $c_{j} = 0$ and $\pi_{j} = 0.001$ for all four of these parameters. This has the desired effect that an apparent 0.005 magnitude transit event that is associated with simultaneous 5 pixel shift away from the star's mean position on the detector would not be considered as a significant event.

The most significant non-linear parameter is the white noise rescaling parameter $r_{\sigma, w}$. In the season-long fit, Eq. \ref{e:loglikelihood} indicates that $P(\mathbb{M}|\mathbb{D})$ would have a shape of
\begin{equation}
\ln P(r_{\sigma, w}|\mathbb{D}) =  - N_{\rm sea} \ln r_{\sigma, w} - \frac{\chi^2_{\rm sea}}{2r_{\sigma, w}^2} 
\end{equation} 
where $\chi^2_{\rm sea}$ is the season-long $\chi^2$ from the $N_{\rm sea}$ observations in the ensemble fit. This is maximized when $\overline{r} = \sqrt{\chi^2_{\rm sea}/N_{\rm sea}}$. We want the nightly prior on $r_{\sigma, w}$ to push it toward $ \overline{r} $, but we also want to provide enough flexibility that nights that are substantially better or worse than typical can be identified as such. To implement this, we mimic the shape of the season-long probability distribution but artificially broaden it with an effective weighting coefficient $N_{\rm eff}$. Propagating this loose prior
\begin{equation}
\ln P(\mathbb{M}) = \cdots - N_{\rm eff} \ln r_{\sigma, w} - \frac{N_{\rm eff}\overline{r}^2}{2r_{\sigma, w}^2} + \cdots
\label{e:prior_r}
\end{equation} 
into the posterior for an individual night, the Maximum A Posteriori (MAP) value of $r_{\sigma, w}$ will be
\begin{equation}
r_{\sigma, w} = \sqrt{\frac{\chi^2 + N_{\rm eff}\overline{r}^2}{N_{\rm obs} + N_{\rm eff}}}.
\label{e:r_MAP}
\end{equation}
We artificially set $N_{\rm eff} = 4$, so on nights with fewer than 4 observations, the MAP value of $r_{\sigma, w}$ will be weighted most toward what the rest of the season says. On nights with more than 4 observations, the data from the night itself will more strongly drive the MAP value. 

We use a modified periodogram \citep{irwin.2011.amefcsrpfmfmts} as part of the season-wide fit to identify the best value of $\overline{v_P}$, the period of the harmonic terms in $V(t)$. We fix $v_{P}$ to this value in all later analysis. While this effectively places an infinitely tight prior on this parameter, the degeneracy between it and the other variability parameters, especially on the timescale of a single night of data, means that its uncertainty is usually accounted for by those terms.

\begin{figure}[tbp]
\begin{center}
\includegraphics[width=\columnwidth]{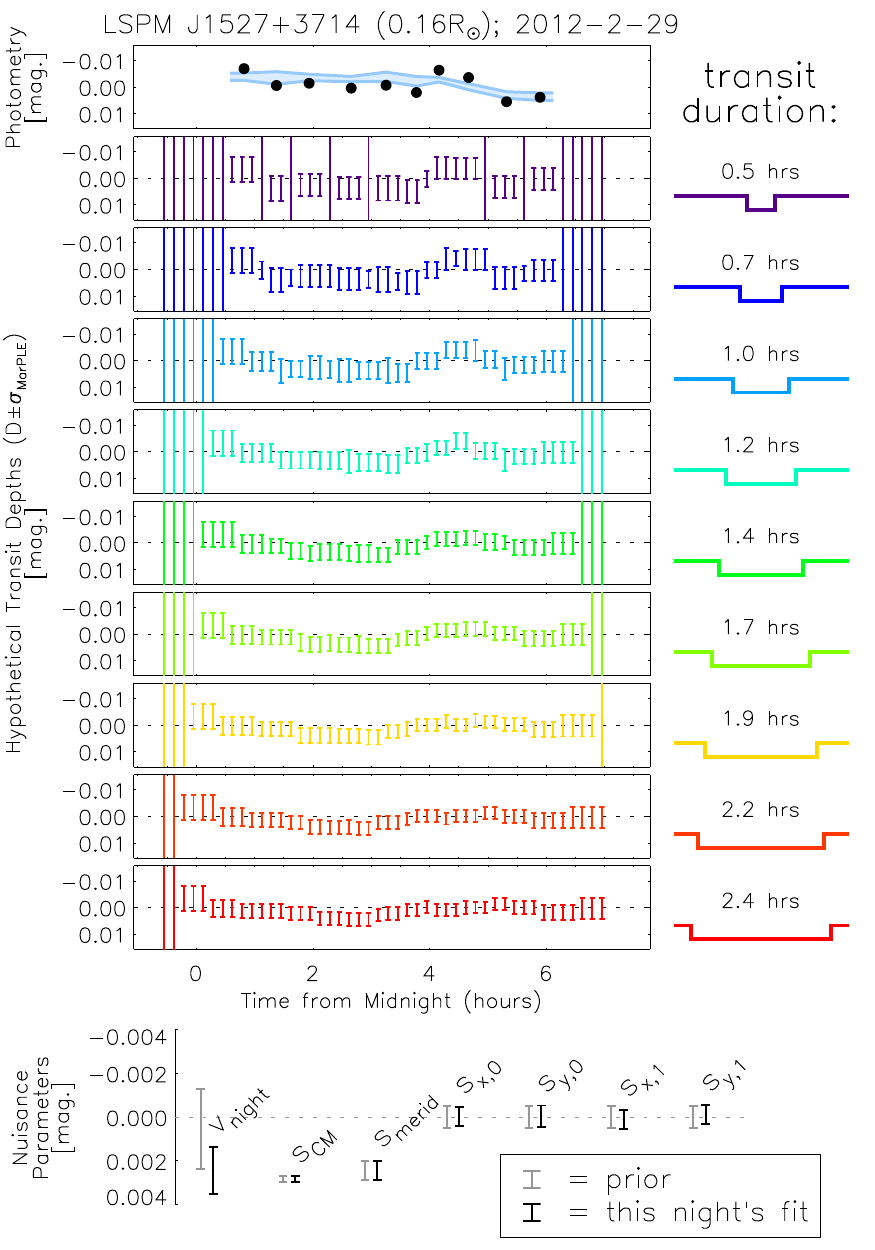}
\caption{\edit{A demonstration of MISS MarPLE applied to one night of MEarth observations, with the star's identifier, inferred stellar radius, and the date indicated above. We show the original MEarth photometry ({\em top panel}, black points), a variability and systematics model with no transits included ({\em same panel}, $\pm1\sigma$; blue swath), and a visualization  ({\em next 9 panels}) of the probability distribution of hypothetical transit depths $P(D | p_{E}, p_{T})$. In this visualization, error bars represent the central $\pm 1\sigma$ confidence regions of the marginalized, Gaussian-shaped $P(D | p_{E}, p_{T})$ for the denoted values of the eclipse epoch $p_E$ (along the time axis) and transit duration $p_T$ (in separate panels). Some of the nuisance variability and systematics parameters over which $\sigma_{\rm MarPLE}$ has been marginalized are also shown ({\em bottom panel}), with the season-long priors in gray and the fits from this night in black. Note that the dip in photometry at the end of the night is not seen as a significant, because it can be explained as a systematic, in this case a variation in the common mode (see the first panel of Figure \ref{f:cmdemo}, from the same night).}}
\label{f:griddemo}
\end{center}
\end{figure}

\subsubsection{Maximizing and Marginalizing}
The shape of $P(\mathbb{M})$ offers a big advantage to our goal of estimating the marginalized transit depth probability distribution. Accounting for all the terms in $P(\mathbb{M|D})$ (Eq. \ref{e:posterior}, \ref{e:loglikelihood}, \ref{e:prior_linear}, and \ref{e:prior_r}), we find that the posterior $P(\mathbb{M}|\mathbb{D})$ can indeed be maximized and marginalized analytically. For fixed $p_E$, and $p_T$, the system of equations
\begin{equation}
\frac{\partial}{\partial c_j}\ln P(\mathbb{M}|\mathbb{D}, p_E, p_T) = \frac{\partial}{\partial c_j}\left( \frac{\chi^2}{r_{\sigma,w}^2} +\Phi^2\right) = 0
\label{e:MAP}
\end{equation}
can be solved exactly for the MAP vector of values $\mathbf{c_{\rm MAP}}$ using only simple matrix operations. The procedure is directly analogous to the problem of weighted linear least squares fitting; see  \citet[][ch. 8]{sivia.2006.dabt} for details of this solution. While not strictly necessary because the priors prevent unconstrained degeneracies in the solution, we use singular value decomposition (SVD) to avoid catastrophic errors in the matrix inversions \citep{press.2002.nrsc}.

The value of $r_{\sigma, w}$ sets the relative weighting between the likelihood and the prior. Thus it is important to estimate $r_{\sigma, w}$ accurately. We solve for it by iterating between Eq. \ref{e:r_MAP} and Eq. \ref{e:MAP}; the solution typically converges to the MAP value within only a couple of iterations. We forego marginalizing over $r_{\sigma, w}$, instead fixing it to its MAP value. Solving for $r_{\sigma, w}$ independently on each night is a better approximation than blindly assuming a global value. 

Importantly, the matrix solution to this problem gives not only the MAP values, it also gives the covariance matrix of the parameters in the fit, which is an exact representation of the shape of $P(\mathbb{M}|\mathbb{D}, p_E, p_T)$, which is a multidimensional Gaussian. The diagonal elements of this covariance matrix give the uncertainty in each parameter {\em marginalized over all the other linear parameters}. Because we have constructed our model in such a way that the parameters that most strongly influence estimates of the transit depth $D$ are linear,  we can use this analytical solution as a robust estimator the shape of the Marginalized Probability of a Lone Eclipse. It gives us both the maximum a posteriori transit depth $\overline{D}$ and the Gaussian width of the distribution $\sigma_{\rm MarPLE}$.

\subsubsection{Are the Priors Really Priors?}
As the priors we use to regularize our model fits are themselves derived from MEarth data, one might object that the division between the likelihood and the prior is set somewhat arbitrarily. We include data only from a single night in the likelihood and group all the information from the rest of the nights into the prior. Indeed, we could have instead organized the entire season of data into the likelihood and left the priors uninformative. The division is arbitrary, but useful.

The advantages of treating nights other than that on which a candidate transit falls as external to likelihood are two-fold. First, it is more computationally efficient: instead of recalculating the likelihood of an entire season's data when investigating individual events, we only need to calculate the likelihood over the relevant night's data points. The information provided by the entire season changes little from candidate transit to candidate transit; thus it is best to store that information as a pre-computed prior. 

Second, this organization scheme allows the flexibility for individual nights to behave differently. For example, consider the pixel position $E_{x,i}(t)$ and $E_{y,i}(t)$ terms in the systematics model, which capture the influence of stars wandering across the detector. As the detector flat-field can change from night to night, it would be foolish to try to fit an entire season's light curve with one set of coefficients for $E_{x,i}(t)$ and $E_{y,i}(t)$; allowing those coefficients to vary from night to night, within a tightly constrained prior, is a more useful approach. Furthermore, dividing the weight of the likelihood and priors as we do provides a helpful degree of outlier resistance, by not forcing the model on any one night to account for strange behavior on one weird night from months before.

\subsubsection{MarPLE in Practice}

We calculate $\overline{D}$ and $\sigma_{\rm MarPLE}$ on a grid of single transit epochs $p_E$ and durations $p_T$. We construct this grid for all nights with usable MEarth data. The epochs in this grid are evenly spaced by $\Delta p_E = 10$ minutes, thus subsampling the typical MEarth observational cadence. The durations are evenly spaced from 0.02 to 0.1 days, spanning the likely durations for the orbital periods to which MEarth has substantial sensitivity. \edit{We extend the grid of $p_E$ before the first and after last observation of each night by half the maximum transit duration, thus probing partial transits.
}

\edit{
We demonstrate this calculation graphically in Fig. \ref{f:griddemo}. First, we show one night of a typical MEarth light curve. To give a sense of a baseline systematics and variability model, we show it the $\pm 1\sigma$ span of model light curves arising from a fit that contains no eclipses. Next, we show a visualization of the MarPLE, the probability distribution of hypothetical eclipse depths $P(D|p_E,p_T)$. For any chosen value of eclipse epoch and duration, the MarPLE is Gaussian-shaped; error bars in Fig. \ref{f:griddemo} show its central $\pm 1 \sigma$ width over the entire grid of $p_E$ and $p_T$. The width of $\sigma_{\rm MarPLE}$ can be seen to decrease for longer durations $p_T$, as more data points are included in each transit window. For epochs and durations with no in-transit points, the transit depth is unconstrained and $\sigma_{\rm MarPLE} \rightarrow \infty$. The transit-like dip in photometry at the end of the night does not register as significant anywhere in the MarPLE. The dip can be explained by MEarth's precipitable water vapor systematic (see first panel of Figure \ref{f:cmdemo}, corresponding to the same night). 
}

\edit{
At the bottom of Fig. \ref{f:griddemo} we also include a subset of the variability and systematics parameters, the ``nuisance parameters'' over which we marginalize. We show error bars representing the Gaussian widths of both the prior $P(\mathbb{M})$, established from the entire season of data, and the results of a fit to this one night of data, $P(\mathbb{M}|\mathbb{D})$. For $v_{\rm night}$, the nightly out-of-transit baseline level parameter, one night's data are more influential than the relatively weak prior, so the fit is notably offset from and tighter than the prior. In contrast, for the remaining nuisance parameters, the influence of one night's data is very weak, so the fit essentially reverts to the input priors. 
}

\edit{
In this example, only 10 data points are contributing to the likelihood. As the model contains almost as many parameters, one might be concerned that we are ``over-fitting'' the data. The bottom of Fig. \ref{f:griddemo} provide an initial step to allay this concern, emphasizing that except for the transit depth, each parameter in the fit has its associated prior that provides its own independent constraint on the parameter. In a pseudo least squares formalism, the presence of these informative priors act as (pseudo) data points, ensuring there are always more ``data'' than parameters. 
}

\edit{
Furthermore and perhaps more importantly, we could indeed be in severe danger of over-fitting if we were interested in the exact values of the cleaned residuals from some single estimate of a best-fit systematics and variability model, but we are safe because we care instead about the marginalized probability of only one particular parameter (the transit depth). Marginalization ignores irrelevant information, so we can include an arbitrary number of nuisance parameters in the fit \citep[see][for discussion]{hogg.2010.darfmd}.  If (and only if) the inferred transit depth at any particular $p_E$ and $p_T$ happens to be strongly covariant with one of these nuisance parameters, then $\sigma_{\rm MarPLE}$ will include a contribution from that parameter. Without the priors degeneracies could potentially inflate $\sigma_{\rm MarPLE}$ to $\infty$, but with the informative season-long priors the nuisance parameters can only vary within the range shown in Fig. \ref{f:griddemo}, limiting the degree to which they can in turn contribute to $\sigma_{\rm MarPLE}$. In the extreme example, if we had a single data point on a night, the cleaned residuals might easily be identically zero (i.e. ``over-fit'') but the MarPLE would accurately express what the night told us about the presence or absence of transits. }
} 

Estimating $\sigma_{\rm MarPLE}$ across the whole grid of $p_E$ for an entire season can be performed very quickly. Each grid point requires only several SVD's of a matrix whose dimension is the sum of the number of data points within the night and the number of linear parameters being fit. For a MEarth light curve containing 1000 points and spanning 100 days, the whole grid of calculations requires several seconds on a typical desktop workstation.

\begin{figure*}[htbp]
\begin{center}
\includegraphics[width=\textwidth]{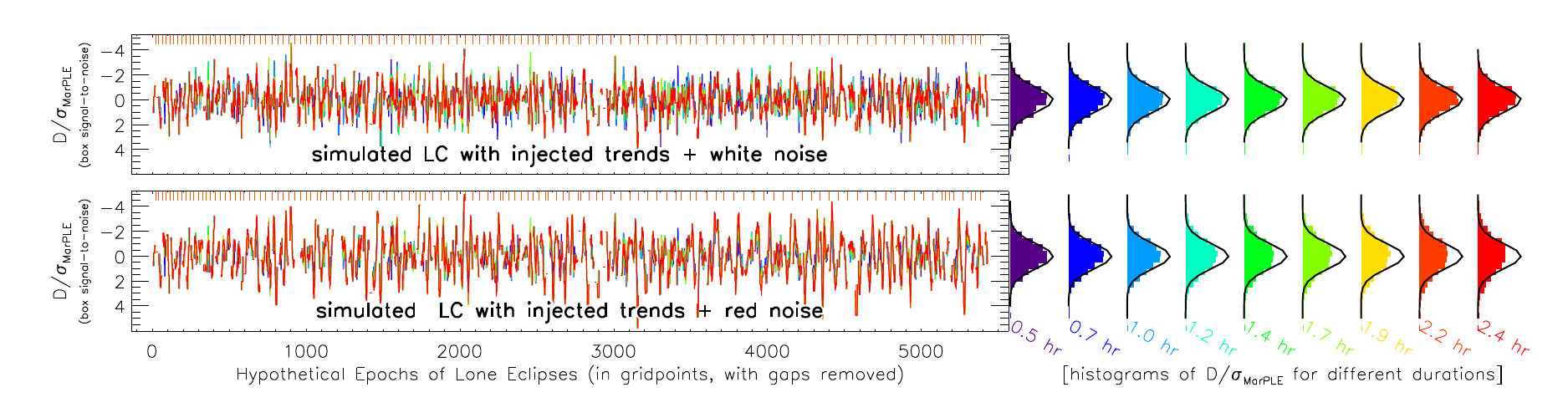}
\caption{A demonstration of the effect of red noise on the inferred significance of transits, showing the signal-to-noise ratio  of hypothetical transits with all possible epochs (along the x-axis) and durations (denoted by color). \edit{$D$, $\sigma_{\rm MarPLE}$, and colors are the same as in Figure \ref{f:griddemo}.} Results are shown for fake light curves generated from the time-stamps of a real MEarth target assuming either white Gaussian noise ({\em top}) or correlated red noise ({\em bottom}), before applying the red noise correction described in \ref{s:rednoise}. Histograms ({\em right}) indicate that uncorrelated white noise leads to $D/\sigma_{\rm MarPLE}$ following a unit-variance Gaussian distribution ({\em black curves}) for all durations, whereas red noise in the light curve broadens the distribution, especially for long duration transits. \edit{If uncorrected, this would cause us to overestimate the significance of candidate transits.}}
\label{f:rednoisedemo}
\end{center}
\end{figure*}

\subsubsection{Ad Hoc Red Noise Correction}\label{s:rednoise}

The likelihood in Eq. \ref{e:loglikelihood} assumed that adjacent light curve data points were statistically independent. If our method fails to completely correct for systematics or stellar variability, this assumption will be violated. Time-correlated noise slows the $\sqrt{N}$ improvement that would be gained by obtaining $N$ independent Gaussian measurements. So, if we were to ignore the temporal correlations between data points, we could substantially bias our estimates of $\sigma_{\rm MarPLE}$. 

Specifically, correlated noise would cause us to overestimate the significance of transits that spanned multiple data points. We demonstrate this phenomenon in Fig. \ref{f:rednoisedemo}, which shows the MarPLE results for two simulated light curves (generated from the real time stamps of a typical MEarth target) with different levels of correlated noise. \edit{One light curve consists of pure white Gaussian noise. The other consists of the white light curve averaged with a smoothed version itself, scaled so both light curves have an identical RMS, roughly approximating a finite red noise contribution. We then inject common mode and meridian flip trends are injected into both light curves. The toy-model simulation of time-correlated noise is very coarse but is only meant to serve an illustrative purpose. }

Because each estimate of $\overline{D}$ is drawn from a Gaussian distribution with a width $\sigma_{\rm MarPLE}$, the quantity $\overline{D}/\sigma_{\rm MarPLE}$ should ideally be Gaussian-distributed around 0 with a variance of 1, except when real transits are present. For the light curve with pure white noise, this is true for all transit durations in Fig. \ref{f:rednoisedemo} (see the histograms at right). For the light curve with significant correlated noise, we underestimate $\sigma_{\rm MarPLE}$ and the distribution of $\overline{D}/\sigma_{\rm MarPLE}$ appears broadened for some durations. The effect is most pronounced at longer durations, where more data points fall within each transit. For the shorter durations, typically only one or two light curve points fall within a transit so the red noise does not substantially affect our estimate of $\sigma_{\rm MarPLE}$. \edit{This general behavior, of overestimating the significance of longer duration transits, is common among MEarth targets whose light curves show features that are poorly matched by the input model.}

To account for the problem, we posit that each light curve has some additional red noise source that can be expressed as a fixed fraction of the white noise, defining $r_{\sigma, r}$ as the ratio of red noise to white noise in a light curve. With this parameterization, the transit depth uncertainty associated with a transit that contains $N_{\rm tra}$ data points becomes
\begin{equation}
\sigma_{\rm MarPLE} = \sigma_{{\rm MarPLE}, w} \times \sqrt{1 + N_{\rm tra} r_{\sigma, r}^2}
\label{e:rednoise}
\end{equation}
where $ \sigma_{{\rm MarPLE}, w}$ is the estimate of $ \sigma_{{\rm MarPLE}}$ that accounted only for white noise. To determine its optimum value, we scale $r_{\sigma, r}$ until the distribution of $\overline{D}/\sigma_{\rm MarPLE}$ has a MAD of $1/1.48$ (i.e. the distribution has a Gaussian width of unity). This correction is similar to the $\mathcal{V}(n)$ formalism described by \citet{pont.2006.enptd}. Henceforth, when we use the term $\sigma_{\rm MarPLE}$, we are referring to its red-noise corrected value.

\begin{figure*}[htbp]
\begin{center}
\includegraphics[width=\textwidth]{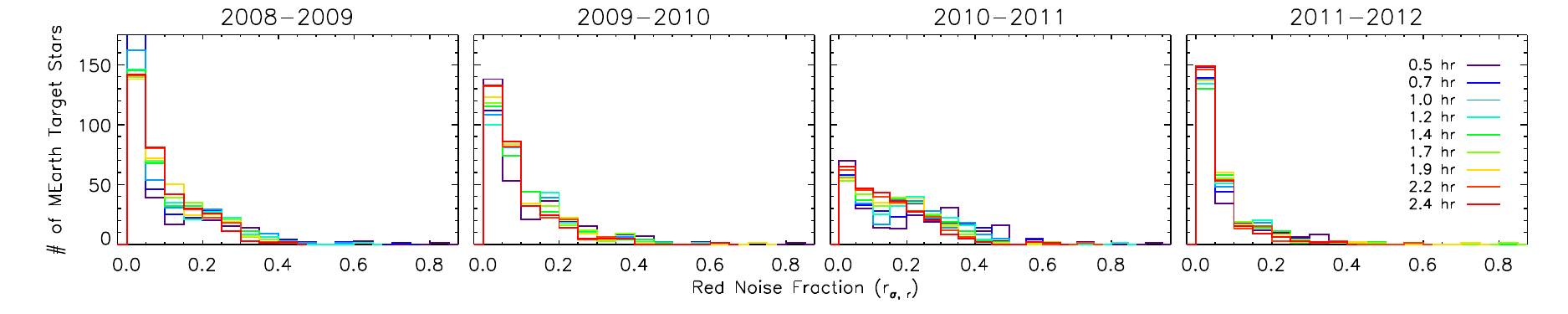}
\caption{Histogram of the red noise rescaling factor $r_{\sigma, r}$ (see Eq. \ref{e:rednoise}) in each of the four MEarth seasons of observations, estimated on different transit duration timescales. We experimented with a narrower filter in the 2010-2011 season in the hopes of alleviating our precipitable water vapor systematic; we found its long wavelength cutoff to be sensitive to humidity and temperature, exacerbating the problem and resulting in increased red noise for this year.}
\label{f:rednoise_summary}
\end{center}
\end{figure*}

A more ideal solution would account for time-correlated noise directly in the likelihood (Eq. \ref{e:loglikelihood}), but doing so would substantially decrease MISS MarPLE's computational efficiency. As such, we settle on Eq. \ref{e:rednoise} as a useful ad hoc solution. Fig. \ref{f:rednoise_summary} shows the amplitude of $r_{\sigma, r}$ for all stars in the MEarth survey, indicating that most stars have low red noise contributions, after accounting for our stellar variability and systematics. 

\subsection{Phasing Multiple MarPLE's Together}\label{s:phasing}

MISS MarPLE, as just described, investigates the significance of a single transit event. The method can be straightforwardly extended to search for periodic transit candidates as well. Once $\overline{D}$ and $\sigma_{\rm MarPLE}$ have been calculated over a grid of $p_E$ and $p_T$, characterizing periodic candidates is simply a matter of combining all precomputed lone eclipses from this grid that match the appropriate period $p_P$ and starting epoch $p_{E0}$. In Kepler parlance, this is the step where Single Event Statistics are combined into Multiple Event Statistics \citep[see][]{tenenbaum.2012.dptsftqkmd}.

Given $p_P$ and $p_{E0}$, we identify those values of $p_E$ that fall within 5 minutes of this linear ephemeris and that have finite values of $\sigma_{\rm MarPLE}$. Each lone eclipse carries its own Gaussian distribution in $D$. Multiplying these independent Gaussians together leads to the standard inverse-variance weighted average:
\begin{equation}
\overline{D_{\rm phased}} = \frac{\sum \overline{D_i}/\sigma_{\rm MarPLE, i}^2}{\sum 1/\sigma_{\rm MarPLE, i}^2}
\label{e:D}
\end{equation}
\begin{equation}
\overline{\sigma_{\rm phased}}^2 =\frac{1}{\sum 1/\sigma_{\rm MarPLE, i}^2}
\label{e:sigma}
\end{equation}
where the sums are performed over the $N_{\rm epoch}$ epochs that were observed for a given candidate period and starting epoch. If $\chi^2_{\rm phased} =  \sum (\overline{D_i} - \overline{D_{\rm phased}})^2/\sigma_{\rm MarPLE, i}^2$ is greater than $N_{\rm epoch}$, we take it as an indication that the uncertainties would have to be underestimated if that candidate ephemeris were real. In this case, we rescale $\overline{\sigma_{\rm phased}}^2$ up by a factor of $\chi^2_{\rm phased}/N_{\rm epoch}$. In other words, we enforce that the independently measured transit depths that go into each phased candidate must agree to within their errors.

\edit{Because we evaluate the MarPLE on grid of epochs that is super-sampled with respect to both our observational cadence (20 minutes) and typical transit durations (0.5 to 2 hours), adjacent values of $p_E$ will have highly correlated transit depth estimates. This simply reflects that a (complicated) binning over the transit duration $p_T$ has already gone into these estimates. Whereas the above sums would be over all in-transit light curve points in a traditional BLS, with MISS MarPLE we include only one term in the sum for each independent event. }

To perform a full search, we repeat this procedure on a grid of periods. Because we hope to identify planets with potentially very few events, it is absolutely crucial that we explore a fine enough grid in periods that we not miss any peaks in the probability distribution. We set $\Delta p_P$ so that when moving from one period to the next, the first and last data points of a season move by 5 minutes with respect to each other in phase (leading to exponentially spaced candidate periods). MEarth target star mass and radius estimates are reliable to 30-35\% \citep[or better for those stars with parallaxes, see][]{nutzman.2008.dcgtshpod}; we use this information to search only up to the transit duration of a  planet in a circular orbit with 0 impact parameter for each period. 

This search is the most computationally intensive step in the process. Searching a typical MEarth season requires roughly $10^5$ candidate periods and 10 minutes on a desktop workstation, using vectorized IDL code. Searching multiple seasons requires up to $10^6$ periods, thus needing correspondingly more time.

\edit{
A brief side note: cleaning methods like TFA or EPD can be run in a ``reconstructive mode," in which they fit away systematics \citep{kovacs.2005.tfawvs} and/or variability \citep{kovacs.2008.atfasms} towards a known signal present in the data. Such reconstructive techniques have generally not been applied when running period searches for planets, because the computational cost of rerunning them for all possible transit periods, durations, and epochs is untenable. When calculating $P(D|p_E, p_T)$ with MISS MarPLE, we are performing an analysis that is in someways similar to a reconstructive TFA/EPD (i.e. fitting systematics and variability in the presence of a candidate transit). But in the case of MISS MarPLE, we first perform this analysis on individual transits using data from individual nights, and phase up the results to candidate periods second. Thus we postpone the combinatorics of the period search until after the costly matrix inversions.
}

The form of the weighted sums in Eq. \ref{e:D} and \ref{e:sigma} highlights an important feature. Events with few observations in a night, events that fall on nights with poor weather, events that correlate with the star's position on the detector, events at high airmass, events on nights where a star is acting weirdly -- namely, bad events -- will have large $\sigma_{\rm MarPLE}$'s and be naturally down-weighted in the sum. In contrast, good events falling on well-sampled, well-behaved nights will get the credit they deserve, exactly as we want. The advantages extend even further, in that this sum can span beyond a single telescope or a single season, enabling the straightforward combination of data from multiple sources with multiple systematics and even at multiple wavelengths into a coherent whole.

\begin{figure*}[htbp]
\begin{center}
\includegraphics[width=0.95\textwidth]{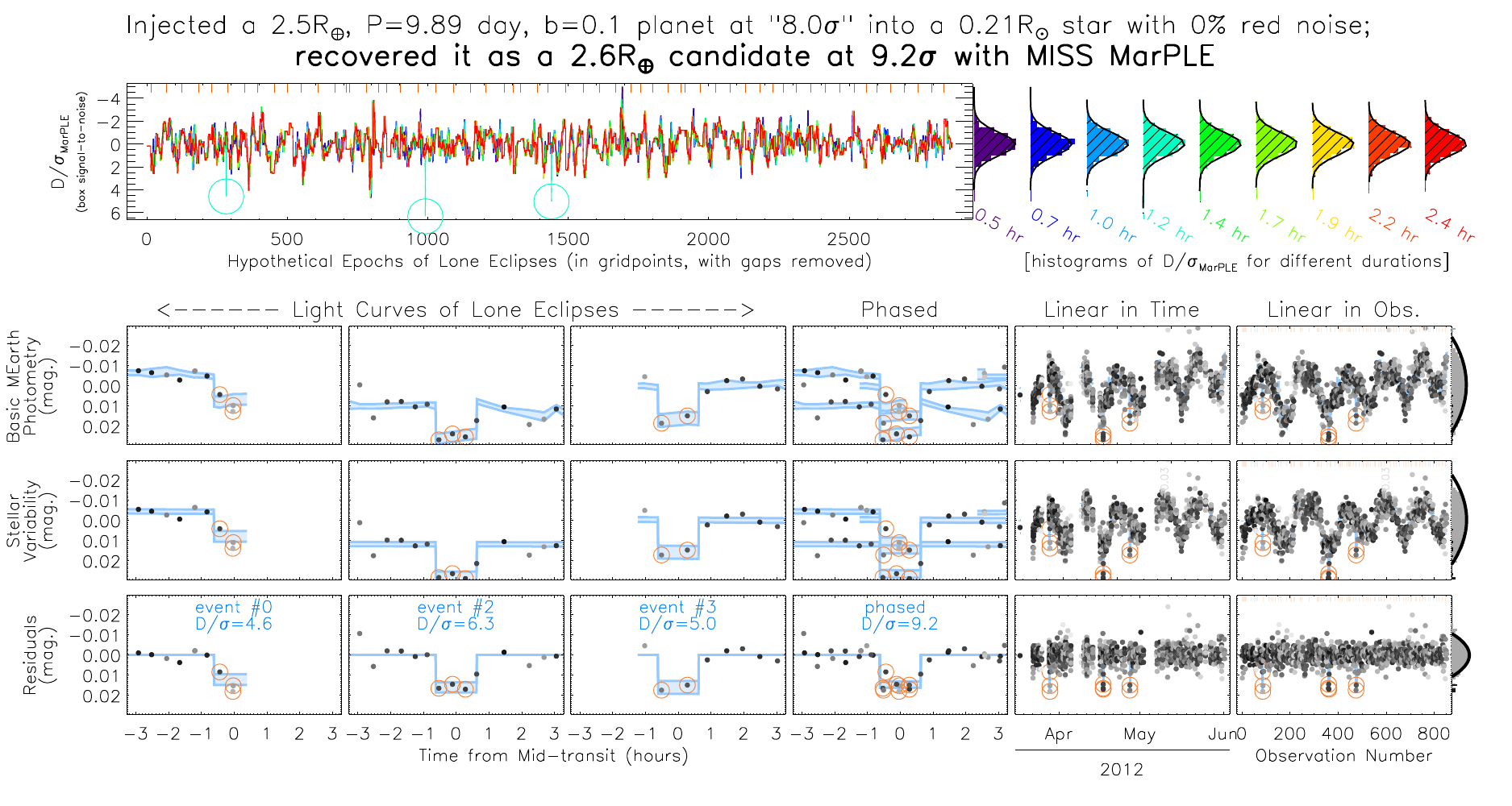}
\caption{An end-to-end demonstration of MISS MarPLE applied to simulated transits injected into a real MEarth light curve. For this candidate, we show $D/\sigma_{\rm MarPLE}$, or the marginalized S/N, for all possible transit epochs and durations ({\em top}), both as an ordered timeseries ({\em left}) and as histograms at fixed duration ({\em right}). We also show MEarth photometry ({\em bottom}, filled circles, with grayscale proportional to $1/\sigma^2$) centered on the individual transit events ({\em \first\--\third\ columns}), phased to the injected planetary period (\fourth\ {\em column}), linearly arranged in time ($5^{th}$ {\em column}), and linearly arranged in observation number ($6^{th}$ {\em column}, with nightly gaps denoted). Light curves are shown for basic MEarth photometry (\first\ {\em row}), after subtracting the systematics model to show stellar variability (\second\ {\em row}), and after subtracting all aspects of the model except for planetary transits (\third\ {\em row}), along with samples from the probability distribution from our light curve model in each panel (blue swaths). Points in-transit are marked throughout this figure.}
\label{f:injectiondemo}
\end{center}
\end{figure*}

\begin{figure*}[htbp]
\begin{center}
\includegraphics[width=0.95\textwidth]{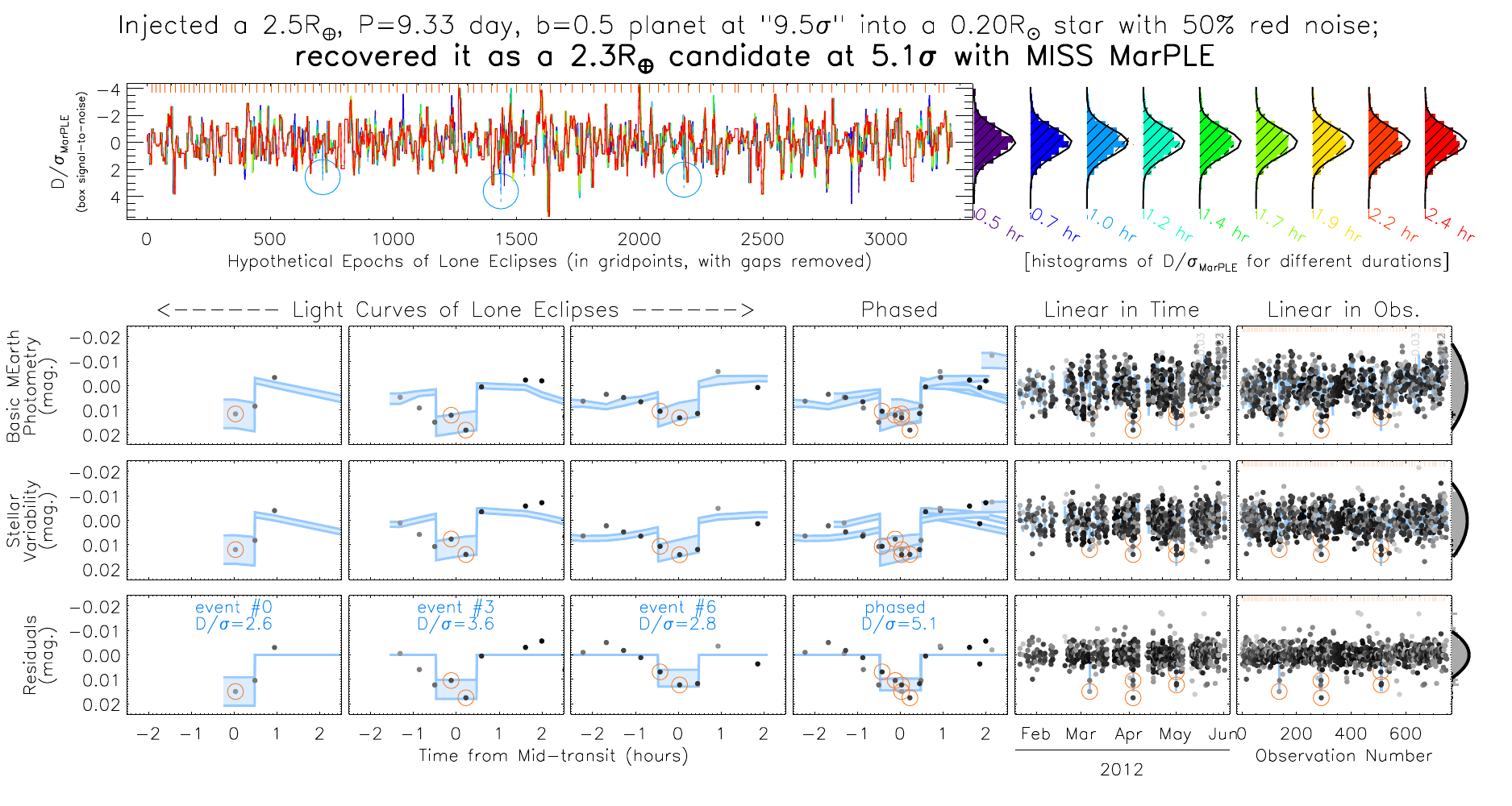}
\caption{Another demonstration as in Figure \ref{f:injectiondemo}, but for a more challenging star. In this case, a high residual red noise fraction and strong covariance between the systematics/variability model and the transit depth limit the recovery significance of this injected candidate. \edit{In a phased search, this candidate would not stand out as strong.}}
\label{f:injectiondemored}
\end{center}
\end{figure*}

\section{Results}\label{s:results}
We apply MISS MarPLE to real MEarth light curves for which we have at least 100 observations in a season, and discuss two aspects of the results here. First, we investigate the properties of simulated transits injected into MEarth light curves, in order to provide concrete examples and compare MISS MarPLE with other methods. Second, we show that the method behaves well when applied to the ensemble of real MEarth light curves and does not generate an overabundance of false positives. 

\edit{Throughout this section, we occasionally point to $D/\sigma_{\rm MarPLE} = 3$ as a characteristic value of interest. MEarth light curves span typically a few thousand independent transit durations, so we expect to find several $3\sigma$ events by chance in each. However, a single candidate event identified at $D/\sigma_{\rm MarPLE} > 3$ significance would be sufficient to set off MEarth's real-time trigger, which would immediately gather new observations to confirm or deny the event. Triggered observations could potentially magnify the significance of the single transit until the chance of it being a false alarm is low: a single transit at $5\sigma$ should formally be expected by chance about once per $3.5\times10^6$ independent epochs tested,  roughly comparable to the number of epochs probed across all the stars in the MEarth survey to date. These thresholds for single events are much lower than that required to eliminate false positives from a phased search for periodic candidates, which as we discuss in Section \ref{s:application} is closer to $7$ or $8\sigma$.}

\subsection{Injected Transits}\label{s:injected}
To show how known transits appear through the lens of MISS MarPLE, we inject simulated transits into each of our raw light curves. Then we apply MISS MarPLE, and compare the significance of the recovered signals to those we injected. For the simulations, we inject 50,000 fake 2-4\rearth\ planets into each MEarth target star, with periods from 0.5 to 20 days, random phases, and impact parameters between 0 and 1. The transits are limb-darkened, using quadratic coefficients for an M4 dwarf \citep{claret.2004.nlsamisfc2tssg}.

We characterize each simulation by an ``injected S/N'': the injected transit depth $D_{\rm injected} = (R_p/R_\star)^2$ divided by $\sigma_{\rm injected}$. We calculate $\sigma_{\rm injected}$ by a $(\sum 1/\sigma^2)^{-1/2}$ estimator, using data points between \second\ and \third\ contact of the injected transit, with a global rescaling to match the RMS of the star's \edit{MAP}-cleaned light curve. In the context of other transit detection algorithms that pair BLS with a pre-search cleaning step, this $D_{\rm injected}/\sigma_{\rm injected}$ has an important meaning. It would be the detection significance BLS would recover for a transit candidate if the pre-search data cleaning perfectly removed variability without influencing the depth of any transit events. Under the assumptions of this idealized BLS, the quantity $D_{\rm injected}/\sigma_{\rm injected}$ is directly linked \citep[see][]{burke.2006.stepssilfswpoc1} to the ``signal residue'' detection statistic in the BLS paper \citep{kovacs.2002.baspt}. 

\subsubsection{Individual Examples}
\edit{We present a couple of illustrative simulations, to give a sense of how MISS MarPLE works. In each case, we use fake planets with three observed transits and periods near ten days. While such long periods would realistically offer this many transits only rarely, we use these hand-picked examples as a convenient way to show both what individual transits of 10 day periods planets look like, and what phasing these transits into periodic candidates looks like. For simplicity's sake, we left MEarth's real-time trigger out of these simulations, showing what individual transits and phased candidates look like in low-cadence data. In reality, most of the injected transits above $3\sigma$ would have been detected by the real-time trigger, and their egresses' populated with additional high-cadence observations. }

Figure \ref{f:injectiondemo} shows one example, a 2.5\rearth\ radius planet with a $P=9.89$ day period and $b=0.1$ impact parameter injected into the raw MEarth light curve of a 0.21\rsun\ star. In this case, the $8.0\sigma$ injected S/N of the transit is well recovered by MISS MarPLE at $9.2\sigma$, as is the inferred planet radius. Three transits fell during times of MEarth observations; they are marked in the plot of $D/\sigma_{\rm MarPLE}$, the eclipse S/N. This star exhibits $0\%$ residual red noise and the transits all fall within well sampled nights; it is thanks to these favorable conditions that the injected and recovered S/N's are so similar. 

For contrast, Figure \ref{f:injectiondemored} shows another example with a different star but broadly similar planetary parameters. Here, the recovered signal's $5.1\sigma$ significance is considerably lower than its injected $9.5\sigma$ strength. One reason for the difference is that the timescale of the intrinsic stellar variability of this star is short enough that the inferred transit depths are substantially correlated with it, thus making a larger contribution to $\sigma_{\rm MarPLE}$. Additionally, our model does not completely remove all the structured features in this light curve so it exhibits a large red noise fraction ($r_{\sigma, r} = 0.5$), further suppressing the detection significance. 

We also show in Figures \ref{f:injectiondemo} and  \ref{f:injectiondemored}  the photometry from MEarth, before and after using the MAP values of our model parameters to subtract off systematics and stellar variability from the light curves. To emphasize that the result of MISS MarPLE is not simply one best-fit model of the systematics and variability, but rather an inferred probability distribution, we plot the swaths of light curve space that are spanned at $\pm1\sigma$ by this distribution of models. We note that the probability distribution $P(\mathbb{M} | p_P, p_{E0}, p_T)$, is conditional on transit period, epoch and duration, so when we visualize the models with the light curves, we have fixed these parameters to their best values (as found in the grid search in Section \ref{s:phasing}). Because the transit search is entangled with the cleaning process, the models and appearance of the MAP-cleaned light curve would be different for different choices of $p_P$, $p_{E0}$, and $p_T$. 

\begin{figure}[tbp]
\begin{center}
\includegraphics[width=\columnwidth]{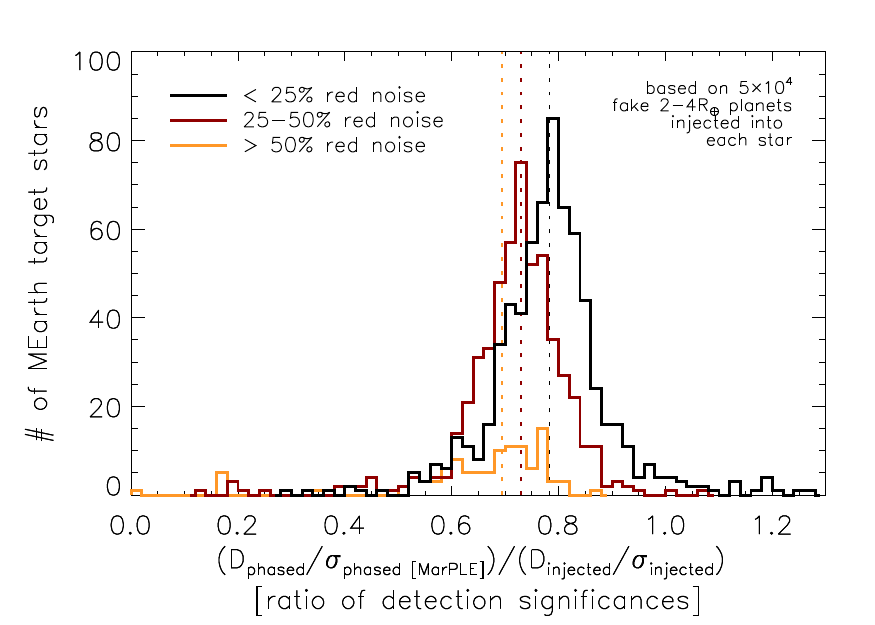}
\caption{A comparison of the significance achieved in a phased search with MISS MarPLE ($\overline{D_{\rm phased}}/\sigma_{\rm phased}$) vs. an idealized BLS ($D_{\rm injected}/\sigma_{\rm injected}$). \edit{ $\overline{D_{\rm phased}}$ and $\sigma_{\rm phased}$ represent the phase-folded combination of in-transit MarPLE's, as in Eq. \ref{e:D} and \ref{e:sigma}. }The definition of $D_{\rm injected}/\sigma_{\rm injected}$ is such that it represents a hypothetical in which any pre-BLS cleaning proceeded perfectly and without influencing the injected transit depth (see text). Each MEarth target star is represented once in this plot by the median of $4\times10^4$ simulations of planets with random periods, phases, impact parameters, and radii. The average significance ratio for each group of residual red noise factors $r_{\sigma, r}$ is shown (dashed lines); as most transits in these simulations contain only 1--2 points the impact of the red noise is relatively muted.}
\label{f:blsfraction}
\end{center}
\end{figure}

\subsubsection{Relationship to BLS}

By itself, a search with BLS will give the significance of a candidate transit that is conditional on the assumption that the out-of-transit baseline flux is constant and that its noise properties are globally known. If preceded by a light curve cleaning step, the transit significance is also conditional on the assumption that the aspects of the cleaning are correct. An important question is how much the marginalized significance of candidate transits found with MISS MarPLE differs from this conditional significance. Generally, the answer to this question will depend on the time sampling of the observations; for a very well-sampled and well-behaved light curve, the BLS and MarPLE results should converge to the same answer. But for the case of the real MEarth data, with its large gaps and fickle systematics, we approach this question with simulations. 

\begin{figure*}[htbp]
\begin{center}
\includegraphics[width=\textwidth]{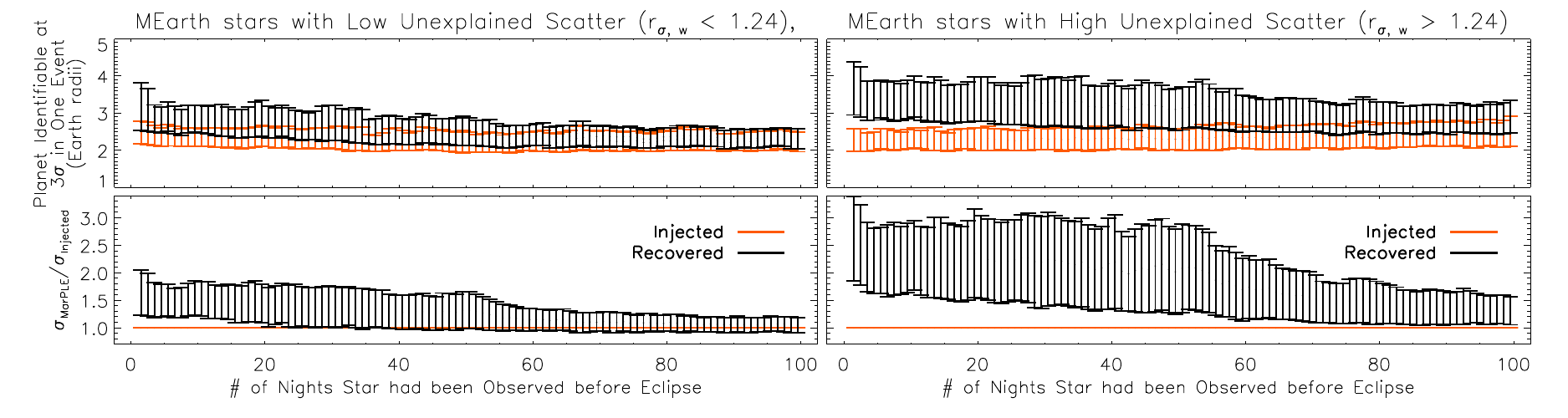}
\caption{The results of ``time-machine'' simulations, in which we inject \edit{single} transits into MEarth light curves and attempt to recover them, using only data {\em up to and including} the transit. As a function of how many nights the target was observed before the candidate transit (and thus how tight the priors can be), we show the smallest planet that could be \edit{identified} at $>3\sigma$ in a single event ({\em top}) and the ratio of the recovered ($\sigma_{\rm MarPLE}$) to injected ($\sigma_{\rm injected}$) transit depth uncertainties ({\em bottom}). Each panel shows results from 50,000 injected transits in each of 100 random stars, with error bars representing the 25\% and 75\% quartiles of the distribution. We show the best ({\em left}) and worst ({\em right}) halves of the MEarth sample, based on how their average white noise rescaling parameter compares to the median of the sample ($r_{\sigma,w} = 1.24$). The lower envelope of each distribution typically corresponds to transits that fall in the middle of well-sampled nights; it converges as soon as tight priors can be established for the systematics coefficients. The upper envelope corresponds more to transits at the starts of nights or in poorly sampled nights; it converges more slowly, depending strongly on the priors for both the systematics and the variability coefficients. }
\label{f:warmup}
\end{center}
\end{figure*}

Figure \ref{f:blsfraction} shows the results of a head-to-head comparison of the significance with which MISS MarPLE views phased (multiple-event) candidates with the significance that would go into a BLS calculation, based on ensemble of injected transits. \edit{A full period search was not run as part of these simulations; we calculated the detection statistics in both cases assuming the period was known. This is in line with our goal with MEarth, that we wish merely to identify whether a signal of a given significance is present, not accurately determine the period of that signal from the existing data.}

For MEarth's best behaved stars (with $r_{\sigma, r} < 0.25$), the marginalized significance estimated by MISS MarPLE is typically $80\%$ of that estimated by our idealized BLS. For these stars, properly accounting for all of the uncertainties in the cleaning process gets us to within $20\%$ of the significance we could achieve in the unrealistic hypothetical that there were no uncertainties in \edit{the cleaning} process. \edit{That MISS MarPLE tends to be more conservative than other methods is very important to our ultimate goal of using candidates identified by MISS MarPLE to invest limited period-finding follow-up observations.} The 20\% factor suppression of transit significance is comparable to the degree to which global filtering methods such as TFA suppress estimated transit depths \citep[e.g. HATNet, see][]{bakos.2012.hgnfaiwt}. However, the advantage of MISS MarPLE is more than simply knowing how much light curve cleaning suppresses transit significance {\em on average}; it is knowing what the cleaning's relative influence is  {\em on individual events} and which events are more, or less, reliable.  MISS MarPLE can give good events on good nights appropriately higher weight, unlike more global methods.

Figure \ref{f:blsfraction} also shows that the penalty imposed by the red noise correction for those stars with $r_{\sigma, r}>0.25$ is significant but not always debilitating. Because MEarth's cadence is so low that typically only 1--2 points fall within any given transit window, the influence of red noise on most transits is relatively small. However, in cases where the cadence is much higher, such as a triggered event observed in real-time with MEarth, the red noise penalty could be much steeper.  Also, as $D_{\rm injected}/\sigma_{\rm injected}$ is the best we could hope to achieve for each candidate, it is an important check that very few stars show significance ratios $>1$.

\subsubsection{Evolution of Priors}

As more nights of observations are gathered, the priors on the systematics and variability parameters associated with a particular star will tighten. As these priors tighten, the significance with which a given transit can be detected will improve. We demonstrate this phenomenon graphically in Figure \ref{f:warmup}, which shows how $\sigma_{\rm MarPLE}$ for single events evolves as more observations are gathered as well as the impact of this evolution on the planet detection. 

We injected transits as before but calculated the MarPLE for every individual event using only the data up to and including the event, excluding all data after \third\ contact. These ``time-machine'' simulations are an approximation to the information available to the MEarth real-time trigger system when deciding whether to gather high-cadence followup of a candidate transit. We show the results for stars in the best and worst halves of the MEarth sample, as judged by how their white noise rescaling factors $r_{\sigma, w}$ compare to the median of the sample $\overline{r_{\sigma, w}}= 1.24$ . Note that transits have a distribution of injected transit depth uncertainties ($\sigma_{\rm injected}$), based on the number of points in transit and the points' relative predicted uncertainties $\sigma_{\rm the}(t)$. 

We highlight in Figure \ref{f:warmup} the smallest planet that could be detected at $3\sigma$ confidence in a single low-cadence event, and how this quantity evolves a function of the number of nights a star is observed before the event. \edit{Imagine a light curve contains 99 event-less nights and one event on the 100th night; Figure \ref{f:warmup} indicates how much the information in the event-less nights improved the reliability of the single event's detection.} In each panel, we show the 25 and 75\% quartiles of the distribution (spanning both multiple stars and multiple random transits). For the stars with low $r_{\sigma, w}$, initially only planets larger than 2.5-3.8\rearth\ exhibit deep enough transits to be detectable. But as more nights of observations tighten the priors, 2.0-2.6\rearth\ planets become detectable, approaching the injected distribution. Stars with high $r_{\sigma, w}$ behave very differently, presumably because our model captures fewer of the features present in the light curves. For these stars, the minimum detectable planet sizes initially span 3.0-4.4\rearth\ and never converge to the injected values. 

We also show in Figure \ref{f:warmup} the distribution of the  ratio $\sigma_{\rm MarPLE}/\sigma_{\rm injected}$ for the simulated transits. The ratio starts off well in excess of unity, but approaches it as more prior-establishing observations are gathered. The range of values it spans corresponds to transits falling at more or less opportune moments. Values of $\sigma_{\rm MarPLE}/\sigma_{\rm injected}$ closer to 1 are usually associated with transits that fall in the middle of a well-behaved night. Higher values correspond to events that fall at the start of a night, events in a night with high excess scatter, or events that coincide with transit-like features in the systematics or variability models.  By the end of a season, the distribution of $\sigma_{\rm MarPLE}/\sigma_{\rm injected}$ for single events in Figure \ref{f:warmup} roughly approaches that for phased candidates in Figure \ref{f:blsfraction}. This makes sense, as the phased S/N ratios in Figure \ref{f:blsfraction} use priors established from all the nights.

\begin{figure*}[htb]
\begin{center}
\includegraphics[width=\textwidth]{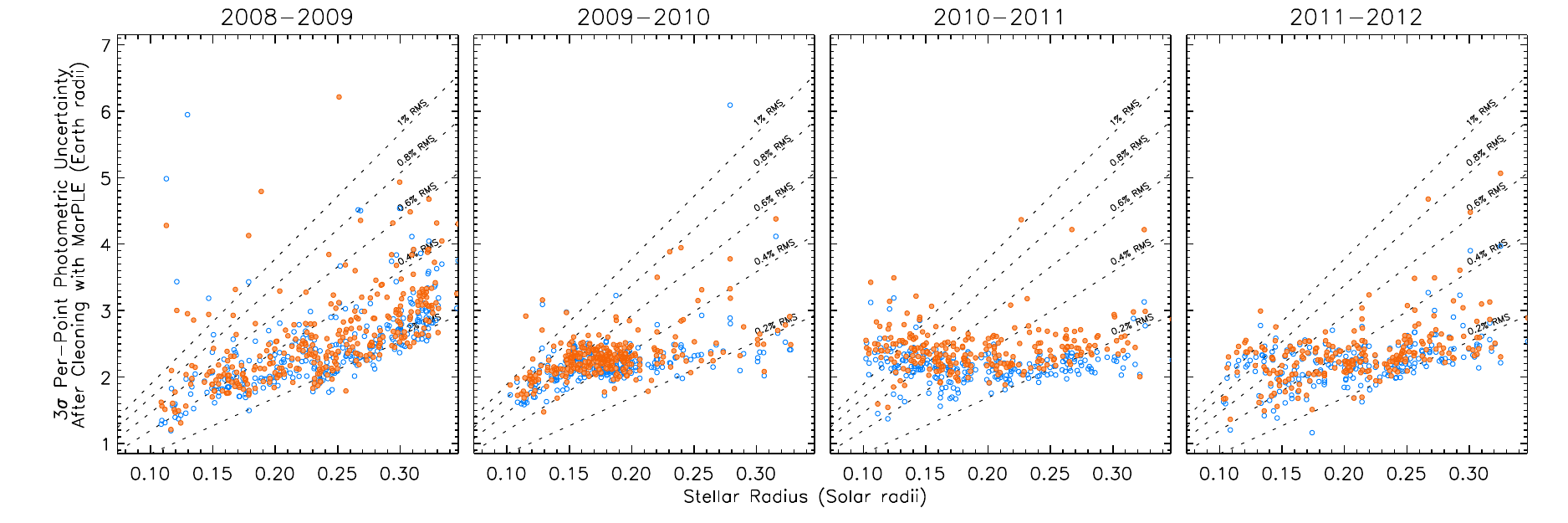}
\caption{The per-point RMS photometric uncertainty as predicted from a CCD noise model ({\em open circles}) and that ultimately achieved in MarPLE-cleaned photometry, after subtracting off MAP models for systematics and stellar variability ({\em filled circles}). In each case, flares and in-transit points for each star's best candidate have been excluded from the calculation of the post-cleaning achieved RMS. Note that the improvement in the RMS relative to Figure \ref{f:rms} is achieved without blind suppression of planetary transits, as the MISS MarPLE cleaning occurs in tandem with the search for transits. }
\label{f:cleanrms}
\end{center}
\end{figure*}

\subsection{Application to MEarth Data}
\label{s:application}
Finally, we employ MISS MarPLE to analyze all the MEarth target stars with no transits injected into them. Figure \ref{f:cleanrms} gives one summary of the method's effectiveness. Here, we plot the achieved RMS in MEarth light curves after using MAP models of the systematics and variability to clean the light curves. Comparison to Figure \ref{f:rms} shows a dramatic improvement, moving the achieved RMS for all the stars much closer to their theoretical minima. However, the achieved RMS values still lie on a locus with a slight upward offset, indicating that our cleaning does not quite reach the photon noise limit. Indeed, this is a reflection of our finding that the median white noise rescaling parameter is $r_{\sigma, w}=1.24$. Figure \ref{f:cleanrms} also shows no evidence that we are over-fitting, in that we never achieve an RMS lower than predicted.

\begin{figure*}[htbp]
\begin{center}
\includegraphics[width=\textwidth]{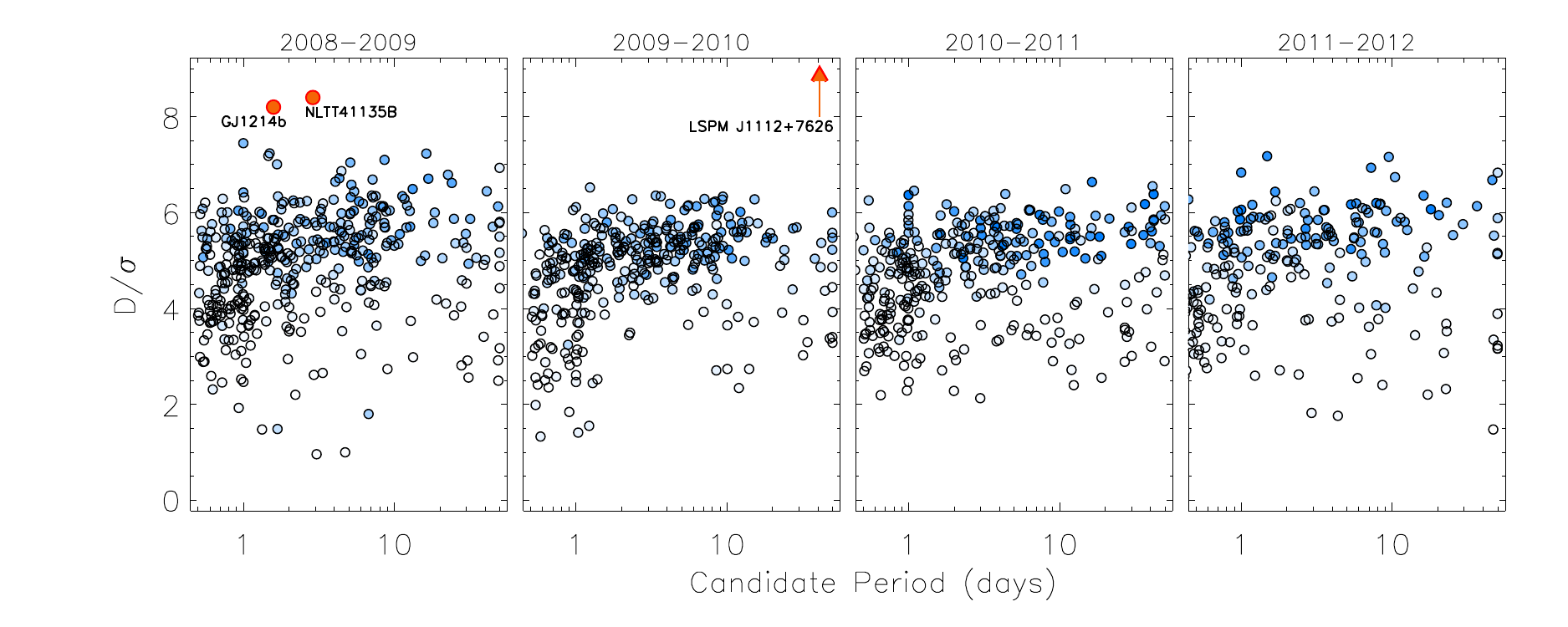}
\caption{A summary of the application of MISS MarPLE to four individual seasons of MEarth data. Each point represents the best periodic candidate identified from a phased search of one star. As each phased transit search effectively performs many effective independent tests on the data, the position of the dense locus of candidates between $5$ and $6\sigma$ is broadly consistent with the null hypothesis, of most of our stars not exhibiting planetary transits. For each star, the intensity of the symbol's color fill is proportional to the number of lone-eclipse epochs ($p_E$) for which observations exist. We also show the detection significance of published MEarth systems, based on the discovery data for each.}
\label{f:candidates}
\end{center}
\end{figure*}

In Figure \ref{f:candidates}, we show the period and detection significance of the best phased candidate that we identify for each MEarth star using our MarPLE-based search. Here we have searched only one season of photometry at a time, so the same star may appear in multiple panels if we had multi-year observations of it. MEarth has published two systems with planet-sized eclipses: the planet GJ1214b \citep{charbonneau.2009.stnls} and the brown-dwarf NLTT41135 \citep[][]{irwin.2010.n4fdbdebtsdmo}. The latter system is in a visual binary that was unresolved in the MEarth discovery data, so its eclipse depth was diluted to a planet-like 2\% depth. While these systems were discovered by using an iterative median-filter \citep{aigrain.2004.ppp} paired with traditional BLS \citep{kovacs.2002.baspt}, we recalculate their detection significances using MISS MarPLE and indicate them in Figure \ref{f:candidates}. We also indicate the long-period low-mass eclipsing binary LSPM J1112+7626 \citep{irwin.2011.ljdmebfmts}, which was detected at very high significance ($30 \sigma$) with the real-time detection trigger. Not shown is the short-period eclipsing binary GJ 3236 \citep{irwin.2009.3bvmebsdmo}, as it was identified by eye in MEarth's commissioning data before the 2008-2009 season. 

\edit{Several new candidates were initially identified above $8\sigma$ significance in Figure \ref{f:candidates}, but  upon inspection the signals were found to be associated with bad raw images. The candidates evaporated after we removed these bad images from consideration.} While we are actively investigating the most promising remaining candidates in Figure \ref{f:candidates}, none are as convincing as were our original confirmed systems \edit{in their discovery data}. 

The morphology of the plots in Figure \ref{f:candidates} is roughly what we expect. Due to geometry, most of our stars will not host exoplanets that transit. Initially, one might think then that the cloud of candidates hovering around $5-6\sigma$ must mean we are substantially overestimating the significance for all of our stars. However, we must consider what makes a reasonable detection threshold for a phased planet search. As discussed in detail by \citet{jenkins.2002.stecpdtp}, each phased search for planets constitutes an enormous number of effective hypotheses being tested against the data. \citet{jenkins.2002.stecpdtp} found that a phased search of a Kepler light curve, with continuous cadence and a 4-year baseline, corresponded to an estimated \edit{number of equivalent independent tests ($N_{\rm EIT}$) of $N_{\rm EIT} = 1.7\times10^7$.}  That is, the detection statistic expected from searching a transit-free Kepler light curve would be the same as asking for the maximum value achieved in $1.7\times10^7$ draws from a unit-variance Gaussian; the median null detection statistic should be above $5\sigma$. It is this consideration that leads to the $7.1\sigma$ detection threshold for the nominal Kepler mission. 

Although the relationship is complicated, generally $N_{\rm EIT}$ increases with the number of observations gathered, the number of periods, and the number of independent phases searched. Because 1-hour transits of M dwarfs are much shorter than the 10-hour transits typical for Kepler, we search many more phases for any given period. While the gap-filled, single-season MEarth light curves going into Figure \ref{f:candidates} have very different properties than Kepler's, an estimate of $N_{\rm EIT}$ on the order of $10^7$ is still a decent estimate. Indeed, using the \citet{jenkins.2002.stecpdtp} bootstrap simulation method, we estimated for a MEarth light curve with $10^3$ data points in which we searched $10^5$ periods that $N_{\rm EIT} \approx 5\times10^6$. Null detection statistics above $5\sigma$ should be a regular occurrence in phased searches of MEarth targets. \edit{The position of MEarth's confirmed targets in Figure \ref{f:candidates} suggests a $7-8\sigma$ threshold is probably appropriate for MEarth}.  \edit{Thresholds could safely be much lower for detecting single events, closer to $5\sigma$, without the brutal combinatorics of a phased search.}

Although it is too computationally intensive to calculate $N_{\rm EIT}$ for the different observational coverages represented by all of the MEarth targets, we try graphically to demonstrate the effect of $N_{\rm EIT}$ in Figure \ref{f:candidates}. We fill the symbols with an intensity proportional to the number of independent eclipse epochs ($p_E$) that the light curve covers, using this as a very rough proxy for $N_{\rm EIT}$. This coloring scheme yields a vertical color gradient in all panels, reflecting the fact that targets with more observations have generally higher $N_{\rm EIT}$ and are more likely to generate high null detection statistics by chance. 

\edit{Time-correlated noise can also disturb the frequency stability of the phased search, if correlations exist over timescales comparable to planetary periods being searched. For example, some uncorrected effect with a $1~{\rm day}^{-1}$ frequency could build up over subsequent nights into what might look like a periodic planet signal. Our ad hoc correction in Section \ref{s:rednoise} does not account for that aspect of time correlated noise. It is likely that such extra uncorrected trends in the 2010-2011 season (which exhibited excess correlated noise, see Table \ref{t:log}) leads to excess of 1 day period candidates in Figure \ref{f:candidates}.}

\section{Future Directions}\label{s:future}
MISS MarPLE could be applied to other ground-based surveys for transiting exoplanets. Its advantages will be greatest for other pointed surveys like MEarth, where individual observations of individual stars are costly enough that it is worth the effort of optimally characterizing the information that each contributes. Aspects of MISS MarPLE be potentially useful to other surveys specifically targeting M dwarfs, such as PTF/M-dwarfs, APACHE, or RoPACS, where the variability and/or systematics are similar to those we described here.

Additionally, ground-based photometric followup to find transits of radial velocity planets \citep[e.g.][]{kane.2009.reetos} faces similar challenges. Typically looking for shallow transits in light curves of bright stars, such efforts require careful consideration of the systematic uncertainties associated with candidate events. For example, the RV-detected super-Earth HD97658b, initially announced to transit from ground-based photoelectric photometry at its predicted time and with $5.7\sigma$ confidence \citep{henry.2011.dtls}, was found not to transit in followup space-based photometry \citep[][]{dragomir.2012.nts9wmp}. This contradiction led to a reevaluation of the systematics in the ground-based observations, which were taken at high airmass. As the most exciting planet discoveries will often be those made very close to the detection threshold, it is important to accurately assess the uncertainties associated with the measured depths of putative transits. Some aspects of a method like the one we proposed here could be useful to marginalize over systematic uncertainties and thus give more confidence in the significance of transit detections in future followup efforts.

Many improvements could be made on our current implementation of MISS MarPLE. For one, the Gaussian likelihood we use to describe our data (Eq. \ref{e:loglikelihood}) is an approximation. It is decent, but it could be elaborated by including a mixture of probability distributions for each data point \citep[to account for junk outliers; e.g.][]{hogg.2010.darfmd,sivia.2006.dabt} or by directly modeling the correlations among data points \citep[see, for example,][]{carter.2009.peftdwcewmatlc}. Also, the variability aspect of our generative light curve model is extremely simplistic (Eq. \ref{e:V(t)}). By replacing our crude sinusoid + nightly offset model with a more sophisticated basis, one might be able to better capture all the variability features in real light curves, thus minimizing the uncertainty its correction injects into the marginalized probability of lone eclipses. In particular, a variability model based on Gaussian processes \citep[see][for an introduction]{gibson.2011.gpfmisats} may be a promising route for setting dynamically evolving priors for the astrophysical behavior of a star on any given night.

\edit{In an upcoming paper, we intend to apply the MISS MarPLE framework to the task of estimating MEarth's sensitivity to 2-4\rearth\ planets over the last four years. Given our single planet detection of GJ1214b, we will use this survey sensitivity estimate to place limits on the occurrence rate of short-period planets around nearby mid-to-late M dwarfs. Such limits would be complementary to results both from Kepler \citep{howard.2012.pow0ssfk} and from the HARPS M dwarf radial velocity program \citep{bonfils.2011.hssepxms}. }

Finally, our ultimate goal with MISS MarPLE is to identify promising candidates with MEarth and make follow-up observations to determine their periods. With this new well-tested method, we plan to focus our efforts in this direction in the years to come. Determining how to schedule the most useful observations for period-finding is a difficult task, but the ``adaptive scheduling'' algorithm proposed by \citet{dzigan.2011.dfslpssteibaas} may prove a very fruitful route. 

\section{Conclusions}\label{s:conclusions}
In this work, we have proposed a new method for detecting planetary transits in wiggly, gap-filled light curves. A method such as this is necessary to eke the optimal sensitivity out of the MEarth Project, our survey for transiting 2-4\rearth\ exoplanets around nearby mid-to-late M dwarfs. MEarth's unique observing strategy gives rise to new challenges (for example, Figure \ref{f:cmdemo}), thus inspiring our efforts to improve on existing transit detection techniques. 

One idea lies at the core of our new method: that when assessing the significance of any individual planetary transit, we want to marginalize over {\em all} the uncertainties, including those associated with cleaning systematics and intrinsic variability from the star's light curve. Our Method for Including Starspots and Systematics in the Marginalized Probability of a Lone Eclipse (MISS MarPLE) can investigate transits within the context of individual nights of observations (see Figure \ref{f:griddemo}), sensibly accounting for various kinds of trends, occasionally messy observational cadences, and the vagaries of photometric conditions common to ground-based observatories. MISS MarPLE uses an analytic, semi-Bayesian approach to include information from an entire season of observations as priors to constrain the expected behavior of a star on any given night. 

We applied MISS MarPLE to four seasons of MEarth photometry, showing that it improves our sensitivity to transiting exoplanets (Figures \ref{f:rms} and \ref{f:cleanrms}). By injecting simulated transiting planets into real MEarth light curves (Figures \ref{f:injectiondemo} and \ref{f:injectiondemored}), we compare MISS MarPLE to the popular Box-fitting Least Squares (BLS) method \citep{kovacs.2002.baspt} and find that even for the best behaved \edit{MEarth targets}, BLS underestimates the true transit depth uncertainties typically by 20\% (Figure \ref{f:blsfraction}). \edit{That is, the covariance of hypothetical transit depths with systematics and variability corrections, on average, increases the true transit depth uncertainty by 20\% for MEarth survey data.} Simulations also show that 2-3\rearth\ planets that are undetectable in the first few weeks a target is observed become detectable, \edit{either in archival data or in incoming data,} later in the season as the behavior of the star is better constrained (Figure \ref{f:warmup}).

The ``MarPLE,'' the probability distribution of hypothetical transit depths for any given transit duration and epoch, is a useful concept. Because this probability distribution is designed to be marginalized over all the complicated factors associated with the telescope or the night on which the observations were taken, it can be treated as a rigorous statistical summary for the presence or absence of a transit at any moment. Thus, we can straightforwardly combine these portable MarPLEs estimated from different telescopes using different filters at different observatories into coherent planet candidates. By properly accounting for so many transit detection uncertainties, the MarPLE should also save precious followup resources by not wasting time on too many false alarms. A framework such as MISS MarPLE could be a useful tool for any collaborative, global followup of long-period transiting exoplanet candidates that may be identified by MEarth or other observatories. 

As the search for transiting planets around nearby stars pushes to radii smaller than 2\rearth, properly accounting for systematics and variability will be become ever more important. MISS MarPLE may prove to be a valuable asset in the hunt for transiting exoplanets around bright M dwarfs in the years to come.

\acknowledgements
We thank Philip Nutzman for inspirational conversations regarding this work; \edit{John Johnson,  Diana Dragomir, Scott Gaudi, and Elisabeth Newton for discussions regarding the method and the paper; and the referee whose careful reading improved the manuscript considerably.} We gratefully acknowledge funding for the MEarth Project from the David and Lucile Packard Fellowship for Science and Engineering and from the National Science Foundation (NSF) under grant number AST-0807690. The MEarth team is greatly indebted to the staff at the Fred Lawrence Whipple Observatory for their efforts in construction and maintenance of the facility and would like to thank Wayne Peters, Ted Groner, Karen Erdman-Myres, Grace Alegria, Rodger Harris, Bob Hutchins, Dave Martina, Dennis Jankovsky, Tom Welsh, Robert Hyne, Mike Calkins, Perry Berlind, and Gil Esquerdo for their support. This research has made use of NASA's Astrophysics Data System.


\begin{thebibliography}{101}
\expandafter\ifx\csname natexlab\endcsname\relax\def\natexlab#1{#1}\fi

\bibitem[{{Aigrain} \& {Favata}(2002)}]{aigrain.2002.bdptmvgmbpsd}
{Aigrain}, S., \& {Favata}, F. 2002, \aap, 395, 625

\bibitem[{{Aigrain} {et~al.}(2004){Aigrain}, {Favata}, \&
  {Gilmore}}]{aigrain.2004.csmpts}
{Aigrain}, S., {Favata}, F., \& {Gilmore}, G. 2004, \aap, 414, 1139

\bibitem[{{Aigrain} {et~al.}(2007){Aigrain}, {Hodgkin}, {Irwin}, {Hebb},
  {Irwin}, {Favata}, {Moraux}, \& {Pont}}]{aigrain.2007.mpsoyoc}
{Aigrain}, S., {Hodgkin}, S., {Irwin}, J.,  {et~al.} 2007, \mnras, 375, 29

\bibitem[{{Aigrain} \& {Irwin}(2004)}]{aigrain.2004.ppp}
{Aigrain}, S., \& {Irwin}, M. 2004, \mnras, 350, 331

\bibitem[{{Alonso} {et~al.}(2004){Alonso}, {Brown}, {Torres}, {Latham},
  {Sozzetti}, {Mandushev}, {Belmonte}, {Charbonneau}, {Deeg}, {Dunham},
  {O'Donovan}, \& {Stefanik}}]{alonso.2004.ttpbs}
{Alonso}, R., {Brown}, T.~M., {Torres}, G.,  {et~al.} 2004, \apjl, 613, L153

\bibitem[{{Bailer-Jones} \& {Lamm}(2003)}]{bailer-jones.2003.lipmbd}
{Bailer-Jones}, C.~A.~L., \& {Lamm}, M. 2003, \mnras, 339, 477

\bibitem[{{Bakos} {et~al.}(2004){Bakos}, {Noyes}, {Kov{\'a}cs}, {Stanek},
  {Sasselov}, \& {Domsa}}]{bakos.2004.wmpwtepd}
{Bakos}, G., {Noyes}, R.~W., {Kov{\'a}cs}, G.,  {et~al.} 2004, \pasp, 116, 266

\bibitem[{{Bakos} {et~al.}(2012){Bakos}, {Csubry}, {Penev}, {Bayliss},
  {Jord{\'a}n}, {Afonso}, {Hartman}, {Henning}, {Kov{\'a}cs}, {Noyes},
  {B{\'e}ky}, {Suc}, {Cs{\'a}k}, {Rabus}, {L{\'a}z{\'a}r}, {Papp}, {S{\'a}ri},
  {Conroy}, {Zhou}, {Sackett}, {Schmidt}, {Mancini}, {Sasselov}, \&
  {Ueltzhoeffer}}]{bakos.2012.hgnfaiwt}
{Bakos}, G.~{\'A}., {Csubry}, Z., {Penev}, K.,  {et~al.} 2012, arXiv:1206.1391

\bibitem[{{Bakos} {et~al.}(2010){Bakos}, {Torres}, {P{\'a}l}, {Hartman},
  {Kov{\'a}cs}, {Noyes}, {Latham}, {Sasselov}, {Sip{\H o}cz}, {Esquerdo},
  {Fischer}, {Johnson}, {Marcy}, {Butler}, {Isaacson}, {Howard}, {Vogt},
  {Kov{\'a}cs}, {Fernandez}, {Mo{\'o}r}, {Stefanik}, {L{\'a}z{\'a}r}, {Papp},
  \& {S{\'a}ri}}]{bakos.2010.hsptbskf}
{Bakos}, G.~{\'A}., {Torres}, G., {P{\'a}l}, A.,  {et~al.} 2010, \apj, 710,
  1724

\bibitem[{{Batalha} {et~al.}(2012){Batalha}, {Rowe}, {Bryson}, {Barclay},
  {Burke}, {Caldwell}, {Christiansen}, {Mullally}, {Thompson}, {Brown},
  {Dupree}, {Fabrycky}, {Ford}, {Fortney}, {Gilliland}, {Isaacson}, {Latham},
  {Marcy}, {Quinn}, {Ragozzine}, {Shporer}, {Borucki}, {Ciardi}, {Gautier},
  {Haas}, {Jenkins}, {Koch}, {Lissauer}, {Rapin}, {Basri}, {Boss}, {Buchhave},
  {Charbonneau}, {Christensen-Dalsgaard}, {Clarke}, {Cochran}, {Demory},
  {Devore}, {Esquerdo}, {Everett}, {Fressin}, {Geary}, {Girouard}, {Gould},
  {Hall}, {Holman}, {Howard}, {Howell}, {Ibrahim}, {Kinemuchi}, {Kjeldsen},
  {Klaus}, {Li}, {Lucas}, {Morris}, {Prsa}, {Quintana}, {Sanderfer},
  {Sasselov}, {Seader}, {Smith}, {Steffen}, {Still}, {Stumpe}, {Tarter},
  {Tenenbaum}, {Torres}, {Twicken}, {Uddin}, {Van Cleve}, {Walkowicz}, \&
  {Welsh}}]{batalha.2012.pcokafmd}
{Batalha}, N.~M., {Rowe}, J.~F., {Bryson}, S.~T.,  {et~al.} 2012,
  arXiv:1202.5852

\bibitem[{{Bean} {et~al.}(2011){Bean}, {D{\'e}sert}, {Kabath}, {Stalder},
  {Seager}, {Miller-Ricci Kempton}, {Berta}, {Homeier}, {Walsh}, \&
  {Seifahrt}}]{bean.2011.ontssgfema}
{Bean}, J.~L., {D{\'e}sert}, J.-M., {Kabath}, P.,  {et~al.} 2011, \apj, 743, 92

\bibitem[{{Bean} {et~al.}(2010){Bean}, {Miller-Ricci Kempton}, \&
  {Homeier}}]{bean.2010.gtsse1}
{Bean}, J.~L., {Miller-Ricci Kempton}, E., \& {Homeier}, D. 2010, \nat, 468,
  669

\bibitem[{{Beatty} {et~al.}(2012){Beatty}, {Pepper}, {Siverd}, {Eastman},
  {Bieryla}, {Latham}, {Buchhave}, {Jensen}, {Manner}, {Stassun}, {Gaudi},
  {Berlind}, {Calkins}, {Collins}, {DePoy}, {Esquerdo}, {Fulton}, {F{\H
  u}r{\'e}sz}, {Geary}, {Gould}, {Hebb}, {Kielkopf}, {Marshall}, {Pogge},
  {Stanek}, {Stefanik}, {Street}, {Szentgyorgyi}, {Trueblood}, {Trueblood}, \&
  {Stutz}}]{beatty.2012.kjtb8psbs}
{Beatty}, T.~G., {Pepper}, J., {Siverd}, R.~J.,  {et~al.} 2012, \apjl, 756, L39

\bibitem[{{Berta} {et~al.}(2011){Berta}, {Charbonneau}, {Bean}, {Irwin},
  {Burke}, {D{\'e}sert}, {Nutzman}, \& {Falco}}]{berta.2011.gsssvtsap}
{Berta}, Z.~K., {Charbonneau}, D., {Bean}, J.,  {et~al.} 2011, \apj, 736, 12

\bibitem[{{Berta} {et~al.}(2012){Berta}, {Charbonneau}, {D{\'e}sert},
  {Miller-Ricci Kempton}, {McCullough}, {Burke}, {Fortney}, {Irwin}, {Nutzman},
  \& {Homeier}}]{berta.2012.ftssgfwfchst}
{Berta}, Z.~K., {Charbonneau}, D., {D{\'e}sert}, J.-M.,  {et~al.} 2012, \apj,
  747, 35

\bibitem[{{Blake} {et~al.}(2008){Blake}, {Bloom}, {Latham}, {Szentgyorgyi},
  {Skrutskie}, {Falco}, \& {Starr}}]{blake.2008.nmudpstc}
{Blake}, C.~H., {Bloom}, J.~S., {Latham}, D.~W.,  {et~al.} 2008, \pasp, 120,
  860

\bibitem[{{Blake} \& {Shaw}(2011)}]{blake.2011.maeugpsr}
{Blake}, C.~H., \& {Shaw}, M.~M. 2011, \pasp, 123, 1302

\bibitem[{{Bonfils} {et~al.}(2011){Bonfils}, {Delfosse}, {Udry}, {Forveille},
  {Mayor}, {Perrier}, {Bouchy}, {Gillon}, {Lovis}, {Pepe}, {Queloz}, {Santos},
  {S{\'e}gransan}, \& {Bertaux}}]{bonfils.2011.hssepxms}
{Bonfils}, X., {Delfosse}, X., {Udry}, S.,  {et~al.} 2011, arXiv:1111.5019

\bibitem[{{Bonfils} {et~al.}(2012){Bonfils}, {Gillon}, {Udry}, {Armstrong},
  {Bouchy}, {Delfosse}, {Forveille}, {Jehin}, {Lendl}, {Lovis}, {Mayor},
  {McCormac}, {Neves}, {Pepe}, {Perrier}, {Pollaco}, {Queloz}, \&
  {Santos}}]{bonfils.2012.utndgdwhvctwtp}
{Bonfils}, X., {Gillon}, M., {Udry}, S.,  {et~al.} 2012, arXiv:1206.5307

\bibitem[{{Bonomo} \& {Lanza}(2008)}]{bonomo.2008.msvdeptiptmhff}
{Bonomo}, A.~S., \& {Lanza}, A.~F. 2008, \aap, 482, 341

\bibitem[{{Bramich} {et~al.}(2005){Bramich}, {Horne}, {Bond}, {Street},
  {Collier Cameron}, {Hood}, {Cooke}, {James}, {Lister}, {Mitchell}, {Pearson},
  {Penny}, {Quirrenbach}, {Safizadeh}, \& {Tsapras}}]{bramich.2005.sptf7}
{Bramich}, D.~M., {Horne}, K., {Bond}, I.~A.,  {et~al.} 2005, \mnras, 359, 1096

\bibitem[{{Burke} {et~al.}(2006){Burke}, {Gaudi}, {DePoy}, \&
  {Pogge}}]{burke.2006.stepssilfswpoc1}
{Burke}, C.~J., {Gaudi}, B.~S., {DePoy}, D.~L., \& {Pogge}, R.~W. 2006, \aj,
  132, 210

\bibitem[{{Carpano} {et~al.}(2003){Carpano}, {Aigrain}, \&
  {Favata}}]{carpano.2003.dptpsvofci}
{Carpano}, S., {Aigrain}, S., \& {Favata}, F. 2003, \aap, 401, 743

\bibitem[{{Carter} \& {Winn}(2009)}]{carter.2009.peftdwcewmatlc}
{Carter}, J.~A., \& {Winn}, J.~N. 2009, \apj, 704, 51

\bibitem[{{Carter} {et~al.}(2011){Carter}, {Winn}, {Holman}, {Fabrycky},
  {Berta}, {Burke}, \& {Nutzman}}]{carter.2011.tlcpxsts1}
{Carter}, J.~A., {Winn}, J.~N., {Holman}, M.~J.,  {et~al.} 2011, \apj, 730, 82

\bibitem[{{Charbonneau} {et~al.}(2009){Charbonneau}, {Berta}, {Irwin}, {Burke},
  {Nutzman}, {Buchhave}, {Lovis}, {Bonfils}, {Latham}, {Udry}, {Murray-Clay},
  {Holman}, {Falco}, {Winn}, {Queloz}, {Pepe}, {Mayor}, {Delfosse}, \&
  {Forveille}}]{charbonneau.2009.stnls}
{Charbonneau}, D., {Berta}, Z.~K., {Irwin}, J.,  {et~al.} 2009, \nat, 462, 891

\bibitem[{{Claret}(2004)}]{claret.2004.nlsamisfc2tssg}
{Claret}, A. 2004, \aap, 428, 1001

\bibitem[{{Clay} {et~al.}(1998){Clay}, {Wild}, {Bird}, {Dawson}, {Johnston},
  {Patrick}, \& {Sewell}}]{clay.1998.cmsrs}
{Clay}, R.~W., {Wild}, N.~R., {Bird}, D.~J.,  {et~al.} 1998, PASA, 15, 332

\bibitem[{{Croll} {et~al.}(2011){Croll}, {Albert}, {Jayawardhana},
  {Miller-Ricci Kempton}, {Fortney}, {Murray}, \&
  {Neilson}}]{croll.2011.btss1smmwa}
{Croll}, B., {Albert}, L., {Jayawardhana}, R.,  {et~al.} 2011, \apj, 736, 78

\bibitem[{{Croll} {et~al.}(2007){Croll}, {Matthews}, {Rowe}, {Gladman},
  {Miller-Ricci}, {Sasselov}, {Walker}, {Kuschnig}, {Lin}, {Guenther},
  {Moffat}, {Rucinski}, \& {Weiss}}]{croll.2007.ls1sstmsp}
{Croll}, B., {Matthews}, J.~M., {Rowe}, J.~F.,  {et~al.} 2007, \apj, 671, 2129

\bibitem[{{Crossfield} {et~al.}(2011){Crossfield}, {Barman}, \&
  {Hansen}}]{crossfield.2011.hdnts1}
{Crossfield}, I.~J.~M., {Barman}, T., \& {Hansen}, B.~M.~S. 2011, \apj, 736,
  132

\bibitem[{{de Mooij} {et~al.}(2012){de Mooij}, {Brogi}, {de Kok},
  {Koppenhoefer}, {Nefs}, {Snellen}, {Greiner}, {Hanse}, {Heinsbroek}, {Lee},
  \& {van der Werf}}]{de-mooij.2012.ontos1wm}
{de Mooij}, E.~J.~W., {Brogi}, M., {de Kok}, R.~J.,  {et~al.} 2012, \aap, 538,
  A46

\bibitem[{{Defa{\"y}} {et~al.}(2001){Defa{\"y}}, {Deleuil}, \&
  {Barge}}]{defay.2001.bmdpt}
{Defa{\"y}}, C., {Deleuil}, M., \& {Barge}, P. 2001, \aap, 365, 330

\bibitem[{{Deming} {et~al.}(2009){Deming}, {Seager}, {Winn}, {Miller-Ricci},
  {Clampin}, {Lindler}, {Greene}, {Charbonneau}, {Laughlin}, {Ricker},
  {Latham}, \& {Ennico}}]{deming.2009.dctseuatsfjwst}
{Deming}, D., {Seager}, S., {Winn}, J.,  {et~al.} 2009, \pasp, 121, 952

\bibitem[{{D{\'e}sert} {et~al.}(2011){D{\'e}sert}, {Bean}, {Miller-Ricci
  Kempton}, {Berta}, {Charbonneau}, {Irwin}, {Fortney}, {Burke}, \&
  {Nutzman}}]{desert.2011.oemasg}
{D{\'e}sert}, J.-M., {Bean}, J., {Miller-Ricci Kempton}, E.,  {et~al.} 2011,
  \apjl, 731, L40+

\bibitem[{{Doyle} {et~al.}(2000){Doyle}, {Deeg}, {Kozhevnikov}, {Oetiker},
  {Mart{\'{\i}}n}, {Blue}, {Rottler}, {Stone}, {Ninkov}, {Jenkins},
  {Schneider}, {Dunham}, {Doyle}, \& {Paleologou}}]{doyle.2000.oltipadsuptmwma}
{Doyle}, L.~R., {Deeg}, H.~J., {Kozhevnikov}, V.~P.,  {et~al.} 2000, \apj, 535,
  338

\bibitem[{{Dragomir} {et~al.}(2012){Dragomir}, {Matthews}, {Howard}, {Antoci},
  {Henry}, {Guenther}, {Kuschnig}, {Marcy}, {Moffat}, {Rowe}, {Rucinski},
  {Sasselov}, \& {Weiss}}]{dragomir.2012.nts9wmp}
{Dragomir}, D., {Matthews}, J.~M., {Howard}, A.~W.,  {et~al.} 2012,
  arXiv:1204.3135

\bibitem[{{Dupuy} \& {Liu}(2009)}]{dupuy.2009.dtjlebsspsd}
{Dupuy}, T.~J., \& {Liu}, M.~C. 2009, \apj, 704, 1519

\bibitem[{{Dzigan} \& {Zucker}(2011)}]{dzigan.2011.dfslpssteibaas}
{Dzigan}, Y., \& {Zucker}, S. 2011, \mnras, 415, 2513

\bibitem[{{Dzigan} \& {Zucker}(2012)}]{dzigan.2012.dtjegpy}
---. 2012, \apjl, 753, L1

\bibitem[{{Evans} {et~al.}(2002){Evans}, {Irwin}, \&
  {Helmer}}]{evans.2002.cmtdss}
{Evans}, D.~W., {Irwin}, M.~J., \& {Helmer}, L. 2002, \aap, 395, 347

\bibitem[{{Fressin} {et~al.}(2012){Fressin}, {Torres}, {Rowe}, {Charbonneau},
  {Rogers}, {Ballard}, {Batalha}, {Borucki}, {Bryson}, {Buchhave}, {Ciardi},
  {D{\'e}sert}, {Dressing}, {Fabrycky}, {Ford}, {Gautier}, {Henze}, {Holman},
  {Howard}, {Howell}, {Jenkins}, {Koch}, {Latham}, {Lissauer}, {Marcy},
  {Quinn}, {Ragozzine}, {Sasselov}, {Seager}, {Barclay}, {Mullally}, {Seader},
  {Still}, {Twicken}, {Thompson}, \& {Uddin}}]{fressin.2012.epok}
{Fressin}, F., {Torres}, G., {Rowe}, J.~F.,  {et~al.} 2012, \nat, 482, 195

\bibitem[{{Fukugita} {et~al.}(1996){Fukugita}, {Ichikawa}, {Gunn}, {Doi},
  {Shimasaku}, \& {Schneider}}]{fukugita.1996.sdsps}
{Fukugita}, M., {Ichikawa}, T., {Gunn}, J.~E.,  {et~al.} 1996, \aj, 111, 1748

\bibitem[{{Giacobbe} {et~al.}(2012){Giacobbe}, {Damasso}, {Sozzetti}, {Toso},
  {Perdoncin}, {Calcidese}, {Bernagozzi}, {Bertolini}, {Lattanzi}, \&
  {Smart}}]{giacobbe.2012.ptspacsfwiaps}
{Giacobbe}, P., {Damasso}, M., {Sozzetti}, A.,  {et~al.} 2012, \mnras, 424,
  3101

\bibitem[{{Gibson} {et~al.}(2011){Gibson}, {Aigrain}, {Roberts}, {Evans},
  {Osborne}, \& {Pont}}]{gibson.2011.gpfmisats}
{Gibson}, N.~P., {Aigrain}, S., {Roberts}, S.,  {et~al.} 2011, arXiv:1109.3251

\bibitem[{{Henry} {et~al.}(2011){Henry}, {Howard}, {Marcy}, {Fischer}, \&
  {Johnson}}]{henry.2011.dtls}
{Henry}, G.~W., {Howard}, A.~W., {Marcy}, G.~W., {Fischer}, D.~A., \&
  {Johnson}, J.~A. 2011, arXiv:1109.2549

\bibitem[{{Hogg} {et~al.}(2010){Hogg}, {Bovy}, \& {Lang}}]{hogg.2010.darfmd}
{Hogg}, D.~W., {Bovy}, J., \& {Lang}, D. 2010, arXiv:1008.4686

\bibitem[{{Howard} {et~al.}(2012){Howard}, {Marcy}, {Bryson}, {Jenkins},
  {Rowe}, {Batalha}, {Borucki}, {Koch}, {Dunham}, {Gautier}, {Van Cleve},
  {Cochran}, {Latham}, {Lissauer}, {Torres}, {Brown}, {Gilliland}, {Buchhave},
  {Caldwell}, {Christensen-Dalsgaard}, {Ciardi}, {Fressin}, {Haas}, {Howell},
  {Kjeldsen}, {Seager}, {Rogers}, {Sasselov}, {Steffen}, {Basri},
  {Charbonneau}, {Christiansen}, {Clarke}, {Dupree}, {Fabrycky}, {Fischer},
  {Ford}, {Fortney}, {Tarter}, {Girouard}, {Holman}, {Johnson}, {Klaus},
  {Machalek}, {Moorhead}, {Morehead}, {Ragozzine}, {Tenenbaum}, {Twicken},
  {Quinn}, {Isaacson}, {Shporer}, {Lucas}, {Walkowicz}, {Welsh}, {Boss},
  {Devore}, {Gould}, {Smith}, {Morris}, {Prsa}, {Morton}, {Still}, {Thompson},
  {Mullally}, {Endl}, \& {MacQueen}}]{howard.2012.pow0ssfk}
{Howard}, A.~W., {Marcy}, G.~W., {Bryson}, S.~T.,  {et~al.} 2012, \apjs, 201,
  15

\bibitem[{{Irwin} {et~al.}(2011{\natexlab{a}}){Irwin}, {Berta}, {Burke},
  {Charbonneau}, {Nutzman}, {West}, \& {Falco}}]{irwin.2011.amefcsrpfmfmts}
{Irwin}, J., {Berta}, Z.~K., {Burke}, C.~J.,  {et~al.} 2011{\natexlab{a}},
  \apj, 727, 56

\bibitem[{{Irwin} {et~al.}(2010){Irwin}, {Buchhave}, {Berta}, {Charbonneau},
  {Latham}, {Burke}, {Esquerdo}, {Everett}, {Holman}, {Nutzman}, {Berlind},
  {Calkins}, {Falco}, {Winn}, {Johnson}, \& {Gazak}}]{irwin.2010.n4fdbdebtsdmo}
{Irwin}, J., {Buchhave}, L., {Berta}, Z.~K.,  {et~al.} 2010, \apj, 718, 1353

\bibitem[{{Irwin} {et~al.}(2009{\natexlab{a}}){Irwin}, {Charbonneau}, {Berta},
  {Quinn}, {Latham}, {Torres}, {Blake}, {Burke}, {Esquerdo}, {F{\"u}r{\'e}sz},
  {Mink}, {Nutzman}, {Szentgyorgyi}, {Calkins}, {Falco}, {Bloom}, \&
  {Starr}}]{irwin.2009.3bvmebsdmo}
{Irwin}, J., {Charbonneau}, D., {Berta}, Z.~K.,  {et~al.} 2009{\natexlab{a}},
  \apj, 701, 1436

\bibitem[{{Irwin} {et~al.}(2009{\natexlab{b}}){Irwin}, {Charbonneau},
  {Nutzman}, \& {Falco}}]{irwin.2009.mpsthsand}
{Irwin}, J., {Charbonneau}, D., {Nutzman}, P., \& {Falco}, E.
  2009{\natexlab{b}}, in IAU Symposium, Vol. 253, IAU Symposium, 37--43

\bibitem[{{Irwin} {et~al.}(2007){Irwin}, {Irwin}, {Aigrain}, {Hodgkin}, {Hebb},
  \& {Moraux}}]{irwin.2007.mpdplcp}
{Irwin}, J., {Irwin}, M., {Aigrain}, S.,  {et~al.} 2007, \mnras, 375, 1449

\bibitem[{{Irwin} {et~al.}(2011{\natexlab{b}}){Irwin}, {Quinn}, {Berta},
  {Latham}, {Torres}, {Burke}, {Charbonneau}, {Dittmann}, {Esquerdo},
  {Stefanik}, {Oksanen}, {Buchhave}, {Nutzman}, {Berlind}, {Calkins}, \&
  {Falco}}]{irwin.2011.ljdmebfmts}
{Irwin}, J.~M., {Quinn}, S.~N., {Berta}, Z.~K.,  {et~al.} 2011{\natexlab{b}},
  \apj, 742, 123

\bibitem[{{Jehin} {et~al.}(2011){Jehin}, {Gillon}, {Queloz}, {Magain},
  {Manfroid}, {Chantry}, {Lendl}, {Hutsem{\'e}kers}, \&
  {Udry}}]{jehin.2011.ttppst}
{Jehin}, E., {Gillon}, M., {Queloz}, D.,  {et~al.} 2011, The Messenger, 145, 2

\bibitem[{{Jenkins}(2002)}]{jenkins.2002.isvdttp}
{Jenkins}, J.~M. 2002, \apj, 575, 493

\bibitem[{{Jenkins} {et~al.}(2002){Jenkins}, {Caldwell}, \&
  {Borucki}}]{jenkins.2002.stecpdtp}
{Jenkins}, J.~M., {Caldwell}, D.~A., \& {Borucki}, W.~J. 2002, \apj, 564, 495

\bibitem[{{Jenkins} {et~al.}(1996){Jenkins}, {Doyle}, \&
  {Cullers}}]{jenkins.1996.mfmgsdtepebad}
{Jenkins}, J.~M., {Doyle}, L.~R., \& {Cullers}, D.~K. 1996, Icarus, 119, 244

\bibitem[{{Kane} {et~al.}(2009){Kane}, {Mahadevan}, {von Braun}, {Laughlin}, \&
  {Ciardi}}]{kane.2009.reetos}
{Kane}, S.~R., {Mahadevan}, S., {von Braun}, K., {Laughlin}, G., \& {Ciardi},
  D.~R. 2009, \pasp, 121, 1386

\bibitem[{{Kim} {et~al.}(2009){Kim}, {Protopapas}, {Alcock}, {Byun}, \&
  {Bianco}}]{kim.2009.dtsavs}
{Kim}, D.-W., {Protopapas}, P., {Alcock}, C., {Byun}, Y.-I., \& {Bianco}, F.~B.
  2009, \mnras, 397, 558

\bibitem[{{Kov{\'a}cs} {et~al.}(2005){Kov{\'a}cs}, {Bakos}, \&
  {Noyes}}]{kovacs.2005.tfawvs}
{Kov{\'a}cs}, G., {Bakos}, G., \& {Noyes}, R.~W. 2005, \mnras, 356, 557

\bibitem[{{Kovacs} \& {Bakos}(2008)}]{kovacs.2008.atfasms}
{Kovacs}, G., \& {Bakos}, G.~A. 2008, Communications in Asteroseismology, 157,
  82

\bibitem[{{Kov{\'a}cs} {et~al.}(2002){Kov{\'a}cs}, {Zucker}, \&
  {Mazeh}}]{kovacs.2002.baspt}
{Kov{\'a}cs}, G., {Zucker}, S., \& {Mazeh}, T. 2002, \aap, 391, 369

\bibitem[{{Kowalski} {et~al.}(2010){Kowalski}, {Hawley}, {Holtzman},
  {Wisniewski}, \& {Hilton}}]{kowalski.2010.wlmds}
{Kowalski}, A.~F., {Hawley}, S.~L., {Holtzman}, J.~A., {Wisniewski}, J.~P., \&
  {Hilton}, E.~J. 2010, \apjl, 714, L98

\bibitem[{{Kundurthy} {et~al.}(2011){Kundurthy}, {Agol}, {Becker}, {Barnes},
  {Williams}, \& {Mukadam}}]{kundurthy.2011.ao1spesa}
{Kundurthy}, P., {Agol}, E., {Becker}, A.~C.,  {et~al.} 2011, \apj, 731, 123

\bibitem[{{Law} {et~al.}(2011){Law}, {Kraus}, {Street}, {Fulton},
  {Hillenbrand}, {Shporer}, {Lister}, {Baranec}, {Bloom}, {Bui}, {Burse},
  {Cenko}, {Das}, {Davis}, {Dekany}, {Filippenko}, {Kasliwal}, {Kulkarni},
  {Nugent}, {Ofek}, {Poznanski}, {Quimby}, {Ramaprakash}, {Riddle},
  {Silverman}, {Sivanandam}, \& {Tendulkar}}]{law.2011.tewmbdstpam}
{Law}, N.~M., {Kraus}, A.~L., {Street}, R.,  {et~al.} 2011, arXiv:1112.1701

\bibitem[{{L{\'e}pine} \& {Shara}(2005)}]{lepine.2005.cnswapmlt0lc}
{L{\'e}pine}, S., \& {Shara}, M.~M. 2005, \aj, 129, 1483

\bibitem[{{McCullough} {et~al.}(2005){McCullough}, {Stys}, {Valenti},
  {Fleming}, {Janes}, \& {Heasley}}]{mccullough.2005.pstepc}
{McCullough}, P.~R., {Stys}, J.~E., {Valenti}, J.~A.,  {et~al.} 2005, \pasp,
  117, 783

\bibitem[{{Menou}(2012)}]{menou.2012.accg}
{Menou}, K. 2012, \apjl, 744, L16

\bibitem[{{Miller} {et~al.}(2008){Miller}, {Irwin}, {Aigrain}, {Hodgkin}, \&
  {Hebb}}]{miller.2008.mpstoc2}
{Miller}, A.~A., {Irwin}, J., {Aigrain}, S., {Hodgkin}, S., \& {Hebb}, L. 2008,
  \mnras, 387, 349

\bibitem[{{Miller-Ricci} \& {Fortney}(2010)}]{miller-ricci.2010.nats1}
{Miller-Ricci}, E., \& {Fortney}, J.~J. 2010, \apjl, 716, L74

\bibitem[{{Miller-Ricci Kempton} {et~al.}(2012){Miller-Ricci Kempton},
  {Zahnle}, \& {Fortney}}]{miller-ricci-kempton.2012.ac1pc}
{Miller-Ricci Kempton}, E., {Zahnle}, K., \& {Fortney}, J.~J. 2012, \apj, 745,
  3

\bibitem[{{Moutou} {et~al.}(2005){Moutou}, {Pont}, {Barge}, {Aigrain},
  {Auvergne}, {Blouin}, {Cautain}, {Erikson}, {Guis}, {Guterman}, {Irwin},
  {Lanza}, {Queloz}, {Rauer}, {Voss}, \& {Zucker}}]{moutou.2005.cbtfptdarslc}
{Moutou}, C., {Pont}, F., {Barge}, P.,  {et~al.} 2005, \aap, 437, 355

\bibitem[{{Naylor} {et~al.}(2002){Naylor}, {Totten}, {Jeffries}, {Pozzo},
  {Devey}, \& {Thompson}}]{naylor.2002.opcda2}
{Naylor}, T., {Totten}, E.~J., {Jeffries}, R.~D.,  {et~al.} 2002, \mnras, 335,
  291

\bibitem[{{Nefs} {et~al.}(2012){Nefs}, {Birkby}, {Snellen}, {Hodgkin},
  {Pinfield}, {Sip{\H o}cz}, {Kovacs}, {Mislis}, {Saglia}, {Koppenhoefer},
  {Cruz}, {Barrado}, {Martin}, {Goulding}, {Stoev}, {Zendejas}, {del Burgo},
  {Cappetta}, \& {Pavlenko}}]{nefs.2012.fuembwts}
{Nefs}, S.~V., {Birkby}, J.~L., {Snellen}, I.~A.~G.,  {et~al.} 2012, \mnras,
  3514

\bibitem[{{Nettelmann} {et~al.}(2011){Nettelmann}, {Fortney}, {Kramm}, \&
  {Redmer}}]{nettelmann.2011.tesmts1}
{Nettelmann}, N., {Fortney}, J.~J., {Kramm}, U., \& {Redmer}, R. 2011, \apj,
  733, 2

\bibitem[{{Nutzman} \& {Charbonneau}(2008)}]{nutzman.2008.dcgtshpod}
{Nutzman}, P., \& {Charbonneau}, D. 2008, \pasp, 120, 317

\bibitem[{{Ofir} {et~al.}(2010){Ofir}, {Alonso}, {Bonomo}, {Carone}, {Carpano},
  {Samuel}, {Weingrill}, {Aigrain}, {Auvergne}, {Baglin}, {Barge}, {Borde},
  {Bouchy}, {Deeg}, {Deleuil}, {Dvorak}, {Erikson}, {Mello}, {Fridlund},
  {Gillon}, {Guillot}, {Hatzes}, {Jorda}, {Lammer}, {Leger}, {Llebaria},
  {Moutou}, {Ollivier}, {P{\"a}etzold}, {Queloz}, {Rauer}, {Rouan},
  {Schneider}, \& {Wuchterl}}]{ofir.2010.sadclcwsusep}
{Ofir}, A., {Alonso}, R., {Bonomo}, A.~S.,  {et~al.} 2010, \mnras, 404, L99

\bibitem[{{Pepper} \& {Gaudi}(2005)}]{pepper.2005.stpss}
{Pepper}, J., \& {Gaudi}, B.~S. 2005, \apj, 631, 581

\bibitem[{{Pollacco} {et~al.}(2006){Pollacco}, {Skillen}, {Collier Cameron},
  {Christian}, {Hellier}, {Irwin}, {Lister}, {Street}, {West}, {Anderson},
  {Clarkson}, {Deeg}, {Enoch}, {Evans}, {Fitzsimmons}, {Haswell}, {Hodgkin},
  {Horne}, {Kane}, {Keenan}, {Maxted}, {Norton}, {Osborne}, {Parley}, {Ryans},
  {Smalley}, {Wheatley}, \& {Wilson}}]{pollacco.2006.wpsc}
{Pollacco}, D.~L., {Skillen}, I., {Collier Cameron}, A.,  {et~al.} 2006, \pasp,
  118, 1407

\bibitem[{{Pont} {et~al.}(2006){Pont}, {Zucker}, \& {Queloz}}]{pont.2006.enptd}
{Pont}, F., {Zucker}, S., \& {Queloz}, D. 2006, \mnras, 373, 231

\bibitem[{{Press}(2002)}]{press.2002.nrsc}
{Press}, W.~H. 2002, {Numerical recipes in C++ : the art of scientific
  computing}, ed. {Press, W.~H.}

\bibitem[{{R{\'e}gulo} {et~al.}(2007){R{\'e}gulo}, {Almenara}, {Alonso},
  {Deeg}, \& {Roca Cort{\'e}s}}]{regulo.2007.twardpt}
{R{\'e}gulo}, C., {Almenara}, J.~M., {Alonso}, R., {Deeg}, H., \& {Roca
  Cort{\'e}s}, T. 2007, \aap, 467, 1345

\bibitem[{{Rogers} \& {Seager}(2010)}]{rogers.2010.tpol1}
{Rogers}, L.~A., \& {Seager}, S. 2010, \apj, 716, 1208

\bibitem[{{Schmidt} {et~al.}(2012){Schmidt}, {Kowalski}, {Hawley}, {Hilton},
  {Wisniewski}, \& {Tofflemire}}]{schmidt.2012.pfaduiel}
{Schmidt}, S.~J., {Kowalski}, A.~F., {Hawley}, S.~L.,  {et~al.} 2012, \apj,
  745, 14

\bibitem[{{Siverd} {et~al.}(2012){Siverd}, {Beatty}, {Pepper}, {Eastman},
  {Collins}, {Bieryla}, {Latham}, {Buchhave}, {Jensen}, {Crepp}, {Street},
  {Stassun}, {Gaudi}, {Berlind}, {Calkins}, {DePoy}, {Esquerdo}, {Fulton},
  {Furesz}, {Geary}, {Gould}, {Hebb}, {Kielkopf}, {Marshall}, {Pogge},
  {Stanek}, {Stefanik}, {Szentgyorgyi}, {Trueblood}, {Trueblood}, {Stutz}, \&
  {van Saders}}]{siverd.2012.ksihispjctms}
{Siverd}, R.~J., {Beatty}, T.~G., {Pepper}, J.,  {et~al.} 2012, arXiv:1206.1635

\bibitem[{Sivia \& Skilling(2006)}]{sivia.2006.dabt}
Sivia, D., \& Skilling, J. 2006, Data Analysis: A Bayesian Tutorial, Oxford
  Science Publications (Oxford University Press)

\bibitem[{{Skrutskie} {et~al.}(2006){Skrutskie}, {Cutri}, {Stiening},
  {Weinberg}, {Schneider}, {Carpenter}, {Beichman}, {Capps}, {Chester},
  {Elias}, {Huchra}, {Liebert}, {Lonsdale}, {Monet}, {Price}, {Seitzer},
  {Jarrett}, {Kirkpatrick}, {Gizis}, {Howard}, {Evans}, {Fowler}, {Fullmer},
  {Hurt}, {Light}, {Kopan}, {Marsh}, {McCallon}, {Tam}, {Van Dyk}, \&
  {Wheelock}}]{skrutskie.2006.ms2}
{Skrutskie}, M.~F., {Cutri}, R.~M., {Stiening}, R.,  {et~al.} 2006, \aj, 131,
  1163

\bibitem[{{Smith} {et~al.}(2012){Smith}, {Stumpe}, {Van Cleve}, {Jenkins},
  {Barclay}, {Fanelli}, {Girouard}, {Kolodziejczak}, {McCauliff}, {Morris}, \&
  {Twicken}}]{smith.2012.kpdcbasec}
{Smith}, J.~C., {Stumpe}, M.~C., {Van Cleve}, J.~E.,  {et~al.} 2012,
  arXiv:1203.1383

\bibitem[{{Street} {et~al.}(2003){Street}, {Horne}, {Lister}, {Penny},
  {Tsapras}, {Quirrenbach}, {Safizadeh}, {Mitchell}, {Cooke}, \& {Collier
  Cameron}}]{street.2003.sptfoc6}
{Street}, R.~A., {Horne}, K., {Lister}, T.~A.,  {et~al.} 2003, \mnras, 340,
  1287

\bibitem[{{Stumpe} {et~al.}(2012){Stumpe}, {Smith}, {Van Cleve}, {Twicken},
  {Barclay}, {Fanelli}, {Girouard}, {Jenkins}, {Kolodziejczak}, {McCauliff}, \&
  {Morris}}]{stumpe.2012.kpdcaaecklc}
{Stumpe}, M.~C., {Smith}, J.~C., {Van Cleve}, J.~E.,  {et~al.} 2012,
  arXiv:1203.1382

\bibitem[{{Tamuz} {et~al.}(2005){Tamuz}, {Mazeh}, \&
  {Zucker}}]{tamuz.2005.cselplc}
{Tamuz}, O., {Mazeh}, T., \& {Zucker}, S. 2005, \mnras, 356, 1466

\bibitem[{{Tenenbaum} {et~al.}(2012){Tenenbaum}, {Christiansen}, {Jenkins},
  {Rowe}, {Seader}, {Caldwell}, {Clarke}, {Li}, {Quintana}, {Smith}, {Stumpe},
  {Thompson}, {Twicken}, {Van Cleve}, {Borucki}, {Cote}, {Haas}, {Sanderfer},
  {Girouard}, {Klaus}, {Middour}, {Wohler}, {Batalha}, {Barclay}, \&
  {Nickerson}}]{tenenbaum.2012.dptsftqkmd}
{Tenenbaum}, P., {Christiansen}, J.~L., {Jenkins}, J.~M.,  {et~al.} 2012,
  \apjs, 199, 24

\bibitem[{{Tingley}(2003)}]{tingley.2003.ietdatc}
{Tingley}, B. 2003, \aap, 408, L5

\bibitem[{{Tingley}(2011)}]{tingley.2011.stdwltbps}
---. 2011, \aap, 529, A6

\bibitem[{{Tofflemire} {et~al.}(2012){Tofflemire}, {Wisniewski}, {Kowalski},
  {Schmidt}, {Kundurthy}, {Hilton}, {Holtzman}, \&
  {Hawley}}]{tofflemire.2012.idfdceiw}
{Tofflemire}, B.~M., {Wisniewski}, J.~P., {Kowalski}, A.~F.,  {et~al.} 2012,
  \aj, 143, 12

\bibitem[{{Tyson} \& {Gal}(1993)}]{tyson.1993.egttfwlfc}
{Tyson}, N.~D., \& {Gal}, R.~R. 1993, \aj, 105, 1206

\bibitem[{{von Braun} {et~al.}(2009){von Braun}, {Kane}, \&
  {Ciardi}}]{von-braun.2009.owfpts}
{von Braun}, K., {Kane}, S.~R., \& {Ciardi}, D.~R. 2009, \apj, 702, 779

\bibitem[{{Young}(1967)}]{young.1967.peavcrts}
{Young}, A.~T. 1967, \aj, 72, 747

\end{thebibliography}
\end{document}